\documentclass[bibyear]{aa}

\usepackage{graphicx}
\usepackage{txfonts}
\usepackage{natbib}
\bibpunct{(}{)}{;}{a}{}{,} 
\usepackage{booktabs}
\usepackage{multicol}
\usepackage{graphicx}
\usepackage{fancyhdr}
\usepackage{amsmath}	
\usepackage{amssymb}	
\usepackage[inter-unit-product=\cdot]{siunitx}
\usepackage{enumerate}
\usepackage{multirow}
\usepackage{mathtools}
\usepackage{longtable}
\usepackage{tabularx}
\usepackage{xcolor} 
\usepackage{ulem} 
\usepackage{comment}
\usepackage{cuted}
\usepackage{soul}
\usepackage{physics}

\usepackage{float}

\usepackage{hyperref}
\hypersetup{colorlinks=true,urlcolor=black, citecolor=blue, linkcolor=black}

\makeatletter
\renewcommand*\aa@pageof{, page \thepage{} of \pageref*{LastPage}}
\makeatother

\newcommand{\orcidicon}[1]{\href{https://orcid.org/#1}{\includegraphics[width=11pt]{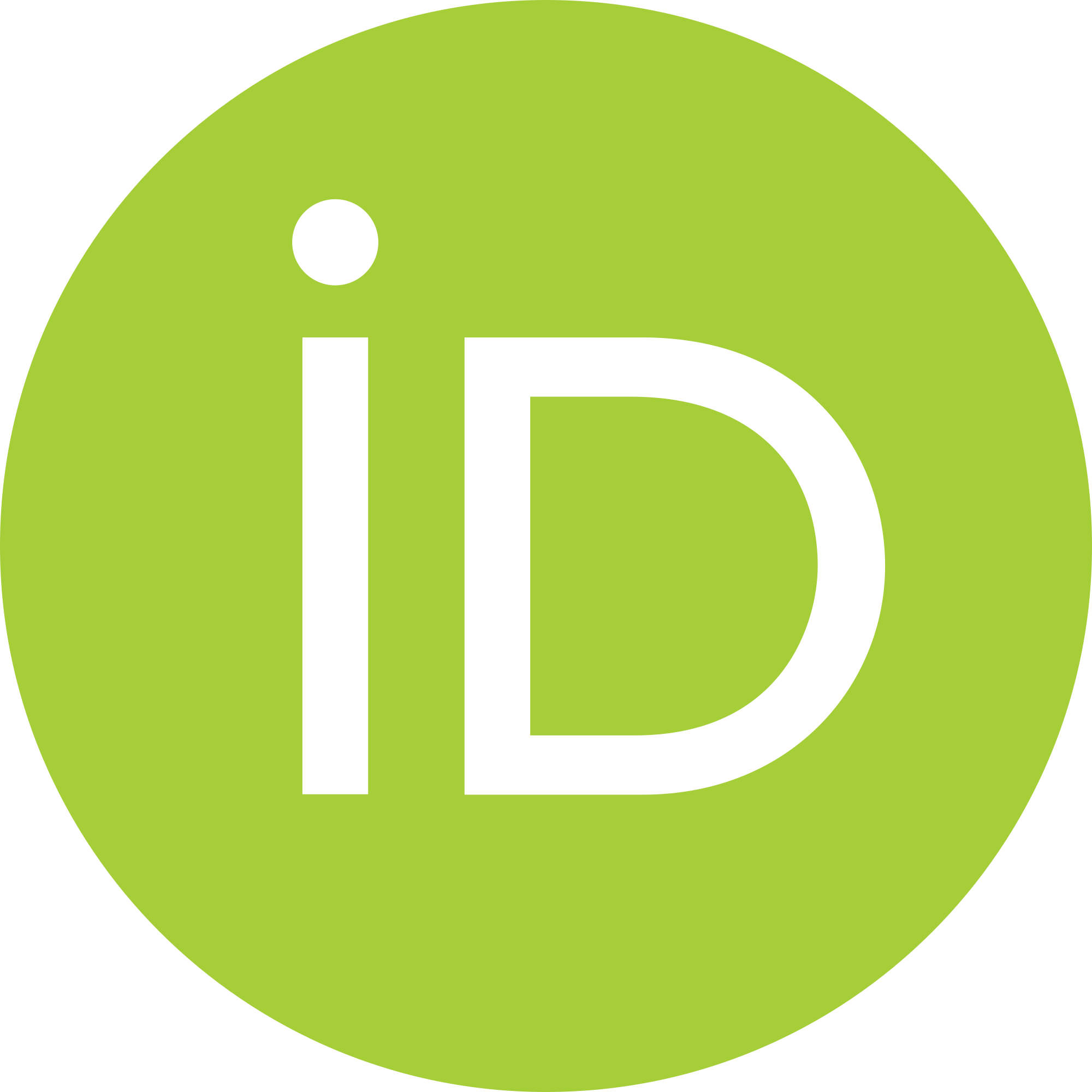}}}
\newcommand{\orcid}[1]{\href{https://orcid.org/#1}{\protect\orcidicon{#1}}}

\newcommand{\msun}{{\rm M}_\odot}

\definecolor{steelblue}{rgb}{0.274 0.510 0.706}

\begin{document}

   \title{Intermediate-mass black hole seeding in \\ galactic nuclei from star cluster migration}
    \titlerunning{Intermediate-mass black hole seeding from star cluster migration}

   \author{
    Stefano Torniamenti\inst{1}
    \orcid{0000-0002-9499-1022} \thanks{\href{mailto:sttorniamenti@mpia.de}{sttorniamenti@mpia.de}}, 
    Nils Hoyer\inst{2} \orcid{0000-0001-8040-4088},
    Nadine Neumayer\inst{1} \orcid{0000-0002-6922-2598},  
    Peter J. Smith\inst{1,3}
    \orcid{0000-0002-7489-5244},
    Manuel Arca Sedda\inst{4,5,6}
    }
    
    \authorrunning{S. Torniamenti et al.}
    \institute{
    $^{1}$Max-Planck-Institut f{\"u}r Astronomie, K{\"o}nigstuhl 17, 69117, Heidelberg, Germany\\
    $^{2}$LIRA, Observatoire de Paris, Universit{\'{e}} PSL, CNRS, Sorbonne Universit{\'{e}}, Universit{\'{e}} Paris Cit{\'{e}}, CY Cergy Paris Universit{\'{e}}, 5 place Jules Janssen, 92195 Meudon, France\\
    $^{3}$Fakult\"at f\"ur Physik und Astronomie, Universit\"at Heidelberg, Im Neuenheimer Feld 226, D-69120, Heidelberg, Germany\\
    $^{4}$Gran Sasso Science Institute, Via F. Crispi 7, L’Aquila, I-67100, Italy \\
    $^{5}$INFN - Laboratori Nazionali del Gran Sasso, I-67100 Assergi, Italy \\
    $^{6}$INAF - Osservatorio Astronomico di Roma, I-00040 Monte Porzio Catone (Rome), Italy
    }
   \date{Received XXXX; accepted YYYY}

 
\abstract{
Nuclear star clusters are one of the most favorable sites to host hierarchical black hole (BH) mergers, potentially bridging the gap from stellar-mass to massive BHs.
However, their assembly and the evolution of their BH populations remain poorly constrained.

We investigate the process of intermediate-mass BH (IMBH) seeding in galactic nuclei from star cluster migration. We introduce \texttt{inSpyral}, a new semi-analytic model that draws star cluster populations from a galaxy formation model (\textit{L-Galaxies 2020}), and integrates their evolution across a wide range of spatial scales, from BH core dynamics to the orbital motion in the host galaxy. 

We find that dynamical friction drives the inspiral of the most massive clusters in galaxies with $M_{\mathrm{\star, gal}} \lesssim 5 \times 10^{10} \,\mathrm{M_\odot}$, seeding their nuclei with IMBHs as early as $z \sim 6$.
The BH mass distribution from BH mergers in migrating clusters extends to  $\sim 300 \, \mathrm{M_\odot}$, a factor of five above the upper limit from in-situ formation. If clusters form with sub-parsec scale radii ($\lesssim 0.5 \, \mathrm{pc}$), hierarchical mergers significantly enhance BH mass growth before migration, and seed galactic nuclei with IMBHs above $10^4 \, \msun$.

The most massive and highly spinning gravitational-wave events are well reproduced by BH mergers involving second-generation remnants that experienced relatively small relativistic kicks ($\lesssim 100 \, \mathrm{km \, s^{-1}}$). GW231123 is consistent with BH mergers between a third-generation primary and a second-generation secondary, which occur in star clusters with mass $> 2 \times 10^{6} \, \msun$.
}


   \maketitle
%


\section{Introduction}

In recent years, observational evidence for the existence of intermediate-mass black holes (IMBHs, $m_{\mathrm{BH}}\sim 10^2$–$10^5 \,\mathrm{M_\odot}$) has steadily increased \citep{greene2020}. 
Observations of fast-moving stars in the core of $\omega$ Centauri have provided the first compelling IMBH candidate in the local Universe ($m_{\mathrm{BH}}> 8\times 10^3 \, \msun$, \citealp{haberle2024}).  
At the lower end of the intermediate-mass regime, gravitational-wave (GW) detections have unveiled candidates with $m_{\mathrm{BH}} \lesssim 300 \, \msun$ \citep{gwtc5pop, gwtc5}. Events such as GW190521 \citep{abbottGW190521,abbottGW190521astro} and GW231123 \citep{gw231123} challenge current models of stellar and binary evolution, with BH components lying within and above the upper-mass gap, where BH formation is strongly suppressed by pair-instability \citep{heger2003,Woosley2007}. The unusually high spins ($\chi \gtrsim 0.7$) of their BH components also point toward a possible hierarchical origin \citep{gerosa2021}.

Hierarchical BH mergers in dense star clusters are one of the most straightforward pathways to overcome the limits of stellar evolution and populate the intermediate-mass regime \citep{miller2002}. In globular clusters, this process can produce a sizable IMBH population with $100-500 \, \msun$ \citep{mapelli2021,arcasedda2021b,arcasedda2026,antonini2023, paiella2025,kiroglu2025b,angeloni2026a,chattopadhyay2026}. Further BH mass growth is quenched by the progressive cluster dissolution \citep{torniamenti2024} and the GW-induced recoil at merger, which in turn depends on the BH spin distribution \citep{campanelli2007}. 

Nuclear star clusters (NSCs, \citealp{genzel2010,georgiev2016,neumayer2020}) are among the most favorable sites for hierarchical mergers . Here, the BH assembly can repeat several times \citep{antonini2016,mapelli2021, arcasedda2023,chattopadhyay2023}, possibly driving the formation of BHs comparable to the IMBH candidates in the nuclei of low-mass galaxies (e.g., NGC 205, \citealp{nguyen2018,nguyen2019}) and $\omega$ Centauri, which is widely believed to be the former nucleus of an accreted galaxy \citep{hilker2000,ibata2019}. 
In NSCs, however, the initial BH mass distribution is highly unconstrained, reflecting significant uncertainties in their formation pathway \citep{neumayer2020}. 

There are two most-debated formation pathways for NSCs. In the in-situ formation scenario, the gas cools and fragments in the central few parsecs of the galactic nucleus, forming stars in a dense NSC \citep{loose1982, milosavljevic2004, walcher2006, carson2015, brown2018, kacharov2018}. In the migration scenario, NSCs are born from the infall of star clusters to the galactic center \citep{tremaine1975,capuzzo2008,antonini2012,arcasedda2014,arcasedda2015}. The aforementioned scenarios are able to individually explain some but not all the observed NSC properties, making it difficult to disentangle a single dominant mechanism \citep{neumayer2020,fahrion2022,arcasedda2023nucl}. 

The NSC formation process has a fundamental impact on the BH population. If stars form in situ, BHs are born in the same environment as their stellar progenitors, possibly from stars with different chemical compositions during multiple formation episodes \citep{seth2006,feldmeierkrause2015, kacharov2018}. If the galactic nucleus is made of migrated star clusters, BHs may have time to form and merge within their parent cluster, potentially seeding the so-formed NSC with IMBHs. In turn, these IMBHs can speed up the formation of massive BH seeds and lead to a non-negligible rate of BBH mergers with peculiar mass ratios and spins \citep{holley-bockelmann2008}.


In this work, we investigate the formation of BH populations in NSCs under the star cluster migration scenario. In particular, we investigate whether this channel can seed galactic nuclei with IMBHs. To this end, we introduce \texttt{inSpyral}, a new semi-analytic model that efficiently captures star cluster evolution across a wide range of spatial scales, from the internal dynamics of BHs in the cluster core to the orbital motion of star clusters within their host galaxies. 
Compared to previous approaches, our model self-consistently combines a semi-analytic treatment of both original and dynamically-formed binary BH (BBH) interactions and mergers with a full integration of the star cluster orbital evolution within arbitrary galactic potentials. We also include a realistic description of the star cluster formation history across cosmic time, by coupling our framework to the galaxy formation model \textit{L-Galaxies 2020} \citep{henriques2020,yates2021,hoyer2025}.

We investigate the conditions that favor or suppress star cluster migration to the galactic center, and explore the physical processes that drive IMBH formation and seeding in galactic nuclei. We account for the major uncertainties related to BH physics (e.g., BH spins), and cluster birth properties (e.g., initial radii). In this way, we provide the first insight into how the NSC formation process shapes IMBH populations in galactic nuclei.

The manuscript is organized as follows. In Sect. \ref{sec:methods} we introduce our new model for evolving BHs and star clusters in galaxies. In Sect. \ref{sec:results}, we show the results for star cluster migration and BH population in galactic nuclei. In Sect. \ref{sec:discussion}, we account for the uncertainties in the process of IMBH seeding. In Sect. \ref{sec:caveats}, we
discuss our results.
Finally, Sect. \ref{sec:concusions} summarizes our conclusions.

\section{Methods} \label{sec:methods}

Over the past decades, semi-analytic codes have emerged as powerful tools for modeling star cluster evolution in galaxies (e.g., see \citealp{gnedin1997,gnedin1999b,chengnedin2022}). Many of these models implement analytic recipes to cosmological simulations, e.g. \textit{E-MOSAICS} \citep{pfeffer2018}, \textit{L-Galaxies} \citep{hoyer2025}, \textit{GAEA} \citep{delucia2024}, to reproduce the formation and evolution of the observed cluster populations. Recently, a new class of semi-analytic code has been developed to investigate BH dynamics in cluster cores. Models like \texttt{cBHBd} \citep{antonini2020, antonini2023, fronimos2026}, \textsc{B-POP} \citep{arcasedda2020, arcasedda2023, arcasedda2026}, \textsc{Fastcluster} \citep{mapelli2021, mapelli2022,torniamenti2024}, and \textsc{Rapster} \citep{kritos2023,kritos2024}, were conceived to efficiently explore the large parameter space involved in the formation of GW events in star clusters.

Here, we present \texttt{inSpyral}, our novel python framework that combines BH dynamical interactions in cluster cores to star cluster evolution and orbit integration in arbitrary galactic tidal fields. By adopting cluster birth properties from \textit{L-Galaxies 2020} \citep{hoyer2025} (Sect. \ref{sec:sc_formation}), our model also incorporates up-to-date prescriptions for cluster formation in galaxies.
In the following, we introduce our formalism for modeling BH populations, including original BBHs (Sect. \ref{sec:bbh_core_init}). Also, we present our approach for describing  cluster evolution (Sect. \ref{sec:sc_evolution}) and integrating their orbits within the host galactic potential (Sect. \ref{sec:sc_in_galaxies}).

\subsection{Galaxy and star cluster formation model} \label{sec:sc_formation}
We describe star cluster formation and galactic potentials using \textit{L-Galaxies 2020} \citep{henriques2020,yates2021,hoyer2025}. This model is built upon merger trees from dark matter–only $N$-body simulations, in which (sub-)halos are identified using the Friends-of-Friends \citep{davis1985a} and Subfind \citep{springel2001c,dolag2009d} algorithms. Here, we consider the realizations run on the Millennium simulations \citep{springel2005c} by \cite{hoyer2025}.
In \textit{L-Galaxies 2020}, the process of cluster formation is implemented through semi-analytic prescriptions, and is calibrated to reproduce observational constraints such as the redshift evolution of the galaxy stellar mass function and the cosmic star formation rate density \citep{henriques2020,yates2021}. We refer to \citet{yates2021} for a detailed description of the 2020 model release, which includes an updated treatment of metal enrichment in the circumgalactic medium.

In this work, we use the galactic potentials and star cluster initial populations presented in \cite{hoyer2025}. 
Each galaxy is resolved into radial annuli to compute the local disk properties that are relevant for star cluster formation, namely the cold gas surface density, epicyclic frequency, and Toomre stability parameter. These quantities are then employed to estimate the cluster formation efficiency \citep[][]{bastian2008b} and the birth properties of star clusters, such as their initial mass and metallicity.

We also account for the clusters that populate galactic halos as a result of galaxy mergers. In \textit{L-Galaxies 2020}, minor mergers occur when the mass ratio between the merging galaxies is $q_{\mathrm{gal}}<0.1$. In this case, the cluster population in the more massive galaxy remains unaltered, while that in the accreted galaxy migrates into the new host. In major mergers, both galactic disks are disrupted, and all star clusters are redistributed into the halo of the merger remnant \citep{hoyer2025}. We refer to \citet{hoyer2025} for a detailed description of the star cluster formation prescriptions and the underlying physical assumptions.

\begin{figure*}
    \centering
    \includegraphics[width=\hsize]{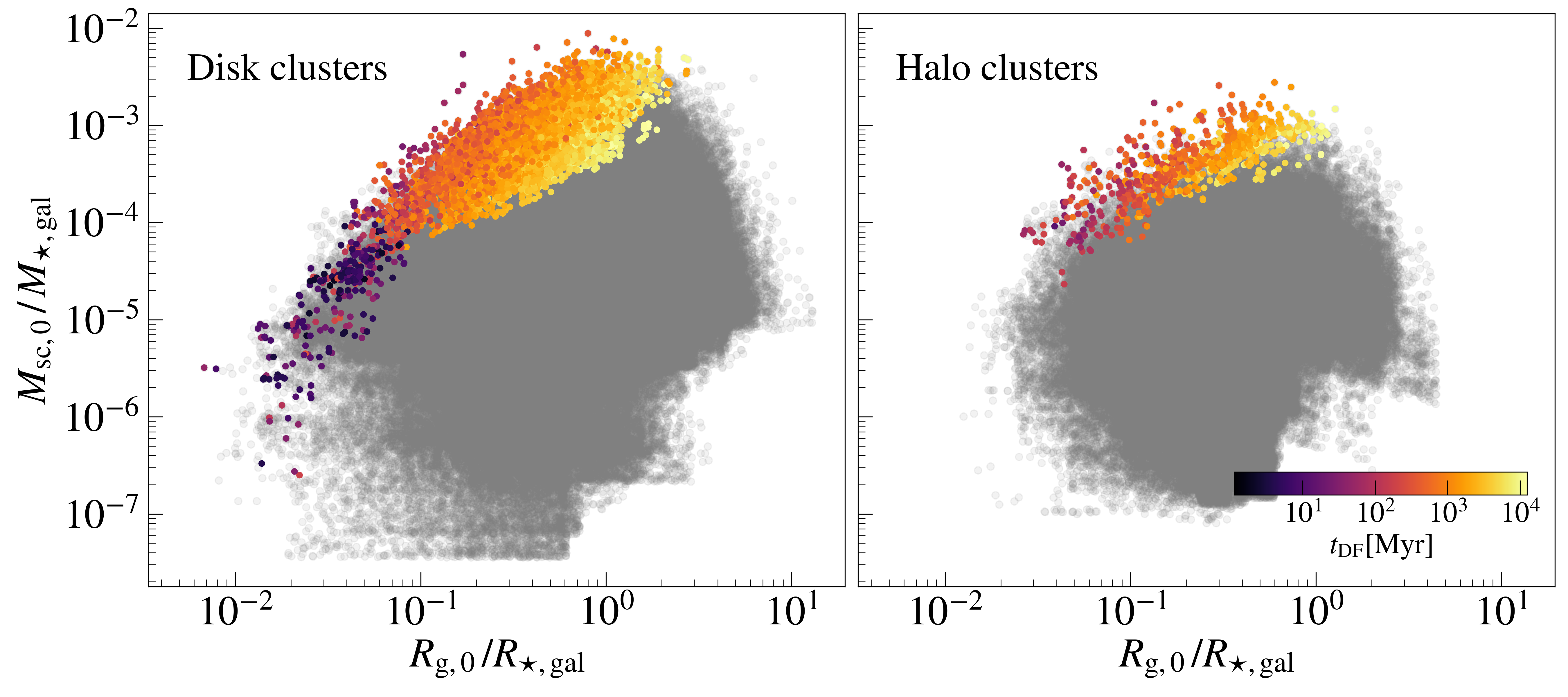}
    \caption{Initial star cluster mass ($M_{\mathrm{sc,0}}$) and galactocentric distance ($R_{\mathrm{g,0}}$) distribution, for all the star clusters (grey points) and for those that migrate to the galactic center (colored points). The quantities have been rescaled by the galactic total stellar mass ($M_{\mathrm{\star, gal}}$) and radial scale ($R_{\mathrm{\star, gal}}$) to enable comparison across different galaxies. For migrating star clusters, the color represents the dynamical-friction time.}
    \label{fig:mcl_rg_scatter}
\end{figure*}

\subsubsection{Galactic potential}
Our galaxy models consist of a stellar, gaseous, and dark matter component. We implement the galactic potentials using the python package for galactic dynamics \texttt{galpy} \citep{bovy2015, webb2019}. For each component, we model the density profile and draw the relevant scale parameters following the analytical approach in \cite{hoyer2025}:
\begin{itemize}
    \item \textbf{Massive BH}. If present, we include the potential of the central massive BH as a point mass at the galactic center:
    \begin{equation}
        \Phi_{\mathrm{BH}}(r) = - \frac{G \, M_{\mathrm{BH}}}{r},
    \end{equation} 
    where $M_{\mathrm{BH}}$ is the central BH mass. \\
    \item \textbf{Galactic bulge}. We model the galactic bulge as a \cite{jaffe1983} sphere:
    \begin{equation}
        \rho_{\mathrm{b}}(r) = \frac{M_{\mathrm{b}}}{4\pi r^3_{\mathrm{b}}}\frac{1}{(r/r_{\mathrm{b}})^2(1+r/r_{\mathrm{b}})},
    \end{equation}
    where $M_{\mathrm{b}}$ is the bulge total mass and $r_{\mathrm{b}}$ the scale radius. \\ 

    \item \textbf{Gas and Stellar disks}. Each galaxy hosts both a gaseous and a stellar disk. We model each disk as the sum of three Miyamoto-Nagai potentials \citep{miyamoto1975}. This implementation delivers an accurate approximation to an exponential disk, up to 10 scale lengths (see \citealp{smith2015}): 
    \begin{equation}
        \rho_{\mathrm{d,i}}(R,z) = \frac{M_{\mathrm{d,i}}}{4 \pi R^2_{\mathrm{d,i}} h_{\mathrm{d,i}}} \exp(-R/R_{\mathrm{d,i}} - |z|/h_{\mathrm{d,i}}).
    \end{equation}
    Here, $M_{\mathrm{d,i}}$ is the total disk component mass, $R_{\mathrm{d,i}}$ and $h_{\mathrm{d,i}}$ are the length the height scales, respectively. We set the disk height scale to $h_{\mathrm{d,i}}=0.1R_{\mathrm{d,i}}$. \\ 
    
    \item \textbf{Dark Matter Halo}. We describe the dark matter halo as a Navarro-Frenk-White potential \citep{navarro1996}:
    \begin{equation}
        \rho_{\mathrm{H}}(r) = \frac{M_{\mathrm{H}}}{4 \pi r^3_{\mathrm{H}}}\frac{1}{(r/r_{\mathrm{H}})(1+r/r_{\mathrm{H}})^2}.
    \end{equation}
    $M_{\mathrm{H}}$ and $r_{\mathrm{H}}$ are the mass and length scales. These quantities are related to the virial mass and radius as $r_{\mathrm{H}}= r_{\mathrm{vir}}/ c_{\mathrm{vir}}$, $M_{\mathrm{H}} = M_{\mathrm{vir}}/ [\log(1+c_{\mathrm{vir}}) - c_{\mathrm{vir}}/(1+c_{\mathrm{vir}})]$, where $c_{\mathrm{vir}}$ is the concentration parameter \citep{dutton2014}. \\ 
    \item \textbf{Hot gas and stellar haloes}. 
    We model the distributions of hot gas and stars from stripped satellite galaxies as isothermal haloes \citep{hoyer2025}:
    \begin{equation}
         \rho_{\mathrm{h,i}}(r) = \frac{M_{\mathrm{h,i}}}{4 \pi r_{\mathrm{h,i}}} \frac{1}{r^2},
    \end{equation}
    where $M_{\mathrm{h,i}}$ is the total halo component mass and $r_{\mathrm{h,i}}=r_{\mathrm{vir}}$.
    
\end{itemize}


\subsubsection{Star cluster birth properties} \label{sec:sc_birth_properties}

We draw the initial star cluster mass ($M_{\mathrm{sc,0}}$), metallicity ($Z$), galactocentric distance ($R_{\mathrm{g,0}}$) and formation redshift from \cite{hoyer2025}. Here, the initial cluster mass is sampled from a truncated power-law distribution: 
\begin{equation}
    \xi (M_{\mathrm{sc,0}}) \propto M^{-2}_{\mathrm{sc,0}} \exp(-M_{\mathrm{sc,0}}/M^{{\mathrm{max}}}_{\mathrm{sc}}),
\end{equation}
where the cutoff mass depends on the local disk properties\footnote{$M^{{\mathrm{max}}}_{\mathrm{sc}}$ is the product of the star formation efficiency, the
bound star formation fraction, the Toomre mass, and the critical fraction of molecular gas to undergo gravitational collapse (Eq.~30 in \citealp{hoyer2025}).}. This initial mass function well reproduces the mass distribution of young star clusters in nearby spiral galaxies \citep{brown2021}, as shown in \cite{hoyer2025}.


We sample the initial half-mass radius using the observation-based approach by \cite{rantala2024}:
\begin{equation} \label{eq:rantala24}
    \frac{r_{\mathrm{h,0}}}{\mathrm{pc}} = \frac{f_{\mathrm{h}} \, R_{\mathrm{4}}}{1.3} \left( \frac{M_{\mathrm{sc,0}}}{10^4 \, \msun} \right)^{\beta},
\end{equation}
where, $\beta=0.180 \pm 0.028$ and $R_{4}=2.365 \pm 0.106$ are inferred from observations of young ($\leq 10 \, \mathrm{Myr}$) star clusters \citep{brown2021} in the LEGUS survey \citep{adamo2017}. 
The factor $f_{\mathrm{h}}= 1/8$ accounts for the smaller radii expected at the time of cluster formation, in agreement with  observations \cite{markskroupa2012} and hydro-dynamical simulations of embedded clusters (\citealp{lahen2025}, see also \citealp{torniamenti2021}). 

The resulting cluster escape velocity is \citep{georgiev2016}:
\begin{equation}
    v_{\mathrm{esc}} = 50 \, f_{\mathrm{c}}  \, \left( \frac{M_{\mathrm{sc}}}{10^5 \, \msun} \right)^{1/3} \left( \frac{\rho_{\mathrm{h}}}{10^5 \, \msun \, \mathrm{pc^{-3}}} \right)^{1/6} \, \mathrm{km \, s^{-1}} ,
\end{equation}
where $\rho_{\mathrm{h}}= 3 M_{\mathrm{sc}} \, / \, (8 \pi r^3_{\mathrm{h}})$ and $f_{\mathrm{c}}=1$, typical of King models with $W_{0}\approx7$. We initialize each cluster with a tangential velocity equal to the local circular velocity, calculated self-consistently from the galactic potential. For disk clusters, we draw the initial vertical velocity component from a Gaussian distribution with a dispersion equal to that of the stellar disk.

\subsubsection{Initial setup}
We draw the star cluster initial conditions and their host galaxy potential from galaxy catalogs at different redshifts. To reconstruct cluster migration across cosmic time, we consider 12 redshift snapshots: $z \simeq 0$, $z \simeq 0.5$, and ten evenly spaced values between $z \simeq 1$ and $10$. At each redshift, we consider 50 different galaxies with stellar mass in the range $M_{\star,\mathrm{gal}} = 10^9 - 6 \times 10^{11} \, \msun$. Each galaxy hosts the disk clusters that have formed since the previous snapshot, as well as those populating the halo as a result of galactic mergers. 

For each galaxy, we consider the 5000 most massive clusters, and select those with initial mass $M_{\mathrm{sc,0}}>10^{4} \, \msun$. Lower-mass clusters are expected to contribute negligibly to BH seeding in the galactic center, due to both low BH retention \citep{pavlik2018} and inefficient cluster migration \citep{arcasedda2014b}. We will confirm this assumption in Sect. \ref{sec:migr_efficiency}. We evolve star clusters in the static galactic potential of their host galaxy at the selected redshift snapshot. In total, our sample consists of $2.4 \times 10^{6}$ star clusters.

\subsection{BH core dynamics} \label{sec:bbh_core_init}

\subsubsection{Initial BH populations: single and original BBHs} \label{sec:bbh_init}

In each star cluster, we model the evolution of the entire BH population. We draw the initial BH population from the catalogs generated with the stellar and binary population-synthesis code \textsc{sevn}  (\citealp{spera2017, Spera2019, Mapelli2020,iorio2023}), 
based on the \textsc{parsec} tracks in \cite{costa2025}. Specifically, we adopt the fiducial model presented in \citet{iorio2023}, which provides (B)BH properties at 14 discrete metallicities in the range $Z \in [10^{-4}, 2\times10^{-2}]${, corresponding to $\mathrm{[Fe/H]} \in [-2.15, +0.15]$.} For each star cluster metallicity, we draw the BH properties from the catalog with the closest metallicity to the nominal value from \textit{L-Galaxies 2020}.

The initial BH population consists of both single BHs and original (i.e., primordial) BBHs. We determine the initial (B)BH number based on the metallicity-dependent formation efficiency by \cite{iorio2023}. For original BBHs, the formation efficiency ranges from $2 \times 10^{-4} \, \msun^{-1}$ at $Z=10^{-4}$ to $4 \times 10^{-5} \, \msun^{-1}$ at $Z=0.02$ (see Fig. 17 in \citealp[]{iorio2023}).

We sample BH natal kicks as \citep{giacobbo2020}:
\begin{equation}\label{eq:SNkick}
    v_{\rm{SN}} = f_{\rm{H05}} \frac{\langle m_{\rm{NS}} \rangle}{m_{\rm{BH}}} \, \frac{m_{\mathrm{ej}}}{\langle m_{\mathrm{ej}} \rangle},
\end{equation}
where $f_{\rm{H05}}$
is a Maxwellian distribution with root-mean square $265 \, \rm{km \,{}s^{-1}}$ (\citealp{hobbs2005}, but see \citealp{disberg2025}). Here, $\langle m_{\rm{NS}} \rangle$ and $\langle m_{\mathrm{ej}} \rangle$ are the average neutron star and ejected mass during the supernova explosion $m_{\mathrm{ej}}$, from \cite{fryer2012}.
If the resulting kick exceeds the cluster escape velocity, we assume that the BH is ejected. We sample the BH dimensionless spin magnitude ($\chi_{\mathrm{BH}}$) from a Maxwellian distribution with $\sigma_{\chi}=0.1$, which is reminiscent of the spins inferred from GW events \citep{gwtc5pop}.

For original BBHs, we draw the initial semi-major axis ($a$) and eccentricity ($e$) from their end state in the \textsc{sevn} simulation. We calculate the tilt angle, with respect to the binary angular momentum, induced by the BH natal kicks following \cite{hurley2002}. If an original BBH is ejected by the supernova kick, we assess whether it merges within a Hubble time\footnote{To calculate the GW merger time with high precision across the whole eccentricity range, we use Eq.~D7 in \cite{iorio2023}.} \citep{peters1964}. Finally, we remove the BBHs that are dynamically soft, as they are easily disrupted by dynamical interactions \citep{heggie1975}.

\begin{figure}
    \centering
    \includegraphics[width=\hsize]{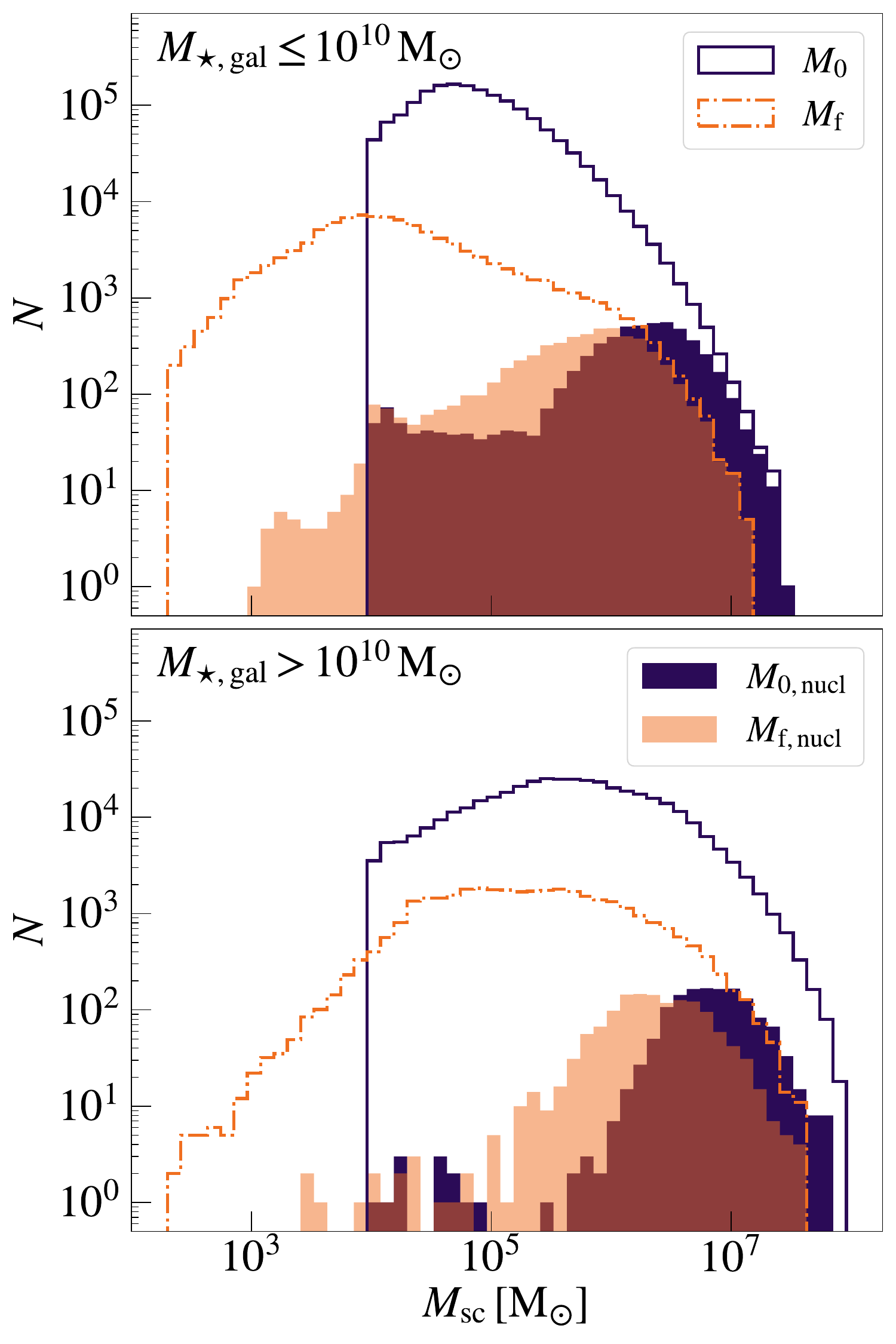}
    \caption{Initial (purple) and final (orange) star cluster mass distributions, for galaxies with stellar mass $M_{\mathrm{\star, gal}} \leq 10^{10} \, \msun$ (upper panel) and $M_{\mathrm{\star, gal}} > 10^{10} \, \msun$ (lower panel). The filled diagrams represent the same distributions for the star clusters that migrate to the galactic center.}
    \label{fig:m0_mf_hist}
\end{figure}

\subsubsection{Dynamical BBHs} \label{sec:dyn_orig_bbh}

After the time of core collapse, star clusters enter a regime of balanced evolution in which dynamical interactions within the core produce energy through BBH formation and hardening \citep{spitzer1987,breen2013a}. The energy production is regulated by the energy demand of the cluster as a whole \citep{henon1961}. We define the onset of balanced evolution as the time of core collapse, $t_{\rm{cc}} \simeq 3 \, t_{\rm{rh, 0}}$, where $t_{\rm{rh, 0}}$ is the initial half-mass relaxation time \citep{antonini2020}. After this, dynamical interactions become effective within the star cluster core, leading to the dynamical formation of BBHs.

We model dynamical BBH formation via three-body interactions following the formalism introduced in \cite{antonini2023}. 
This criterion preferentially pairs the most massive BHs, and is based on the hard binaries formation rate \citep{heggie1975}:
\begin{equation} \label{eq:Gamma}
    \Gamma_{\mathrm{3b}}(m_1,m_2,m_3) \propto  \frac{n(m_1) \,{}n(m_2) \,{}n(m_3)\,{}m_1^4\,{} m_2^4\,{} m_3^{5/2}}{{(m_1+m_2+m_3)^{1/2}(m_1+m_2)^{1/2}}} \,{}\beta^{9/2},
\end{equation}
where $n_{i}=n(m_i)$ is the number density of BHs with mass $m_i$ (with $i=1,$ 2, and 3) and $\beta{}$ is the inverse of their mean internal energy. We assume that $\beta{}$ is constant, as expected in the case of energy equipartition among the most massive objects in the cluster center \citep{spitzer1987}.
{For each star cluster, we compute $n(m_{i})$} as the BH mass function of single BHs in the core. We derive the probability density function for sampling $m_{1,2}$ by integrating Eq.~\ref{eq:Gamma} over $m_3$. We use this probability density function to draw $m_1$ and $m_2$, and select the two BHs with the closest masses as the dynamical BBH components (see, e.g., \citealp{torniamenti2024}).

We generate the initial BBH eccentricity from a thermal distribution  
and the semi-major axis at the hard-soft limit \citep{heggie1975}. 
We assume that the spins are isotropically oriented over a sphere, since dynamical encounters remove any alignment with the binary angular momentum \citep{rodriguez2015}.


\begin{figure}
    \centering
    \includegraphics[width=\hsize]{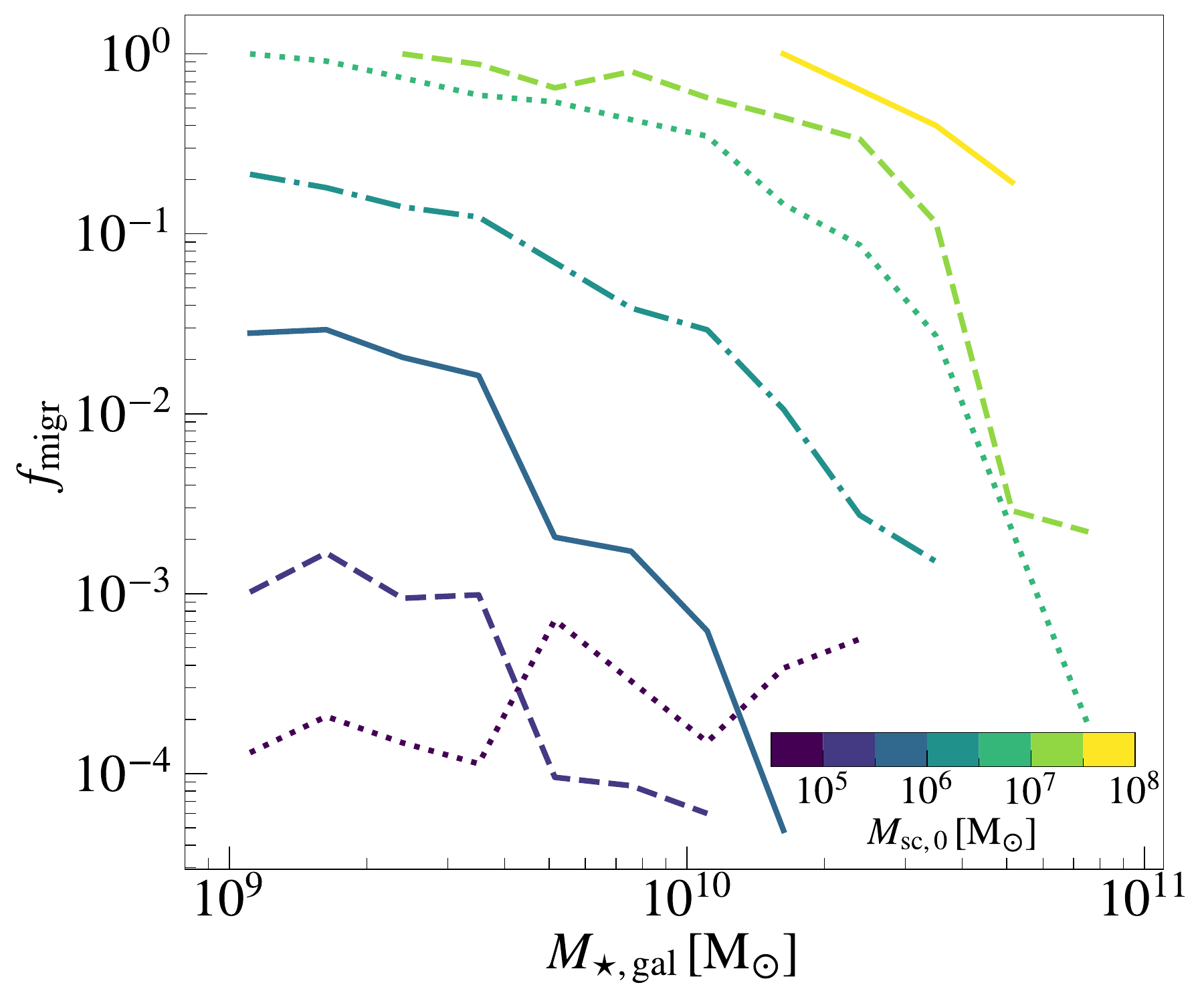}
    \caption{Fraction of migrating star clusters ($f_{\mathrm{migr}}$) in galaxies with different stellar mass. $f_{\mathrm{migr}}$ approaches unity for the most massive clusters that form in a galaxy and decreases with increasing galaxy mass.
    }
    \label{fig:migr_efficiency}
\end{figure}

\subsubsection{BBH hardening} \label{sec:bh_bal_ev}
In a state of balanced evolution, energy within the core is primarily generated by the hardening of a single BBH per time \citep{breen2013a}\footnote{This assumption may not be valid if primordial binaries are present (see, e.g., \citealp{trenti2006}). In Appendix \ref{app:internal_evolution}, we test our model on $N-$body simulations with non-zero binary fractions.}, and is then transferred to the stellar bulk of the cluster \citep{henon1961}. To identify the BBH that is responsible for energy production, we first sample a dynamical BBH following the procedure described in Sect. \ref{sec:dyn_orig_bbh}. We then assign an interaction rate to each original BBH \citep{hills1980}:
\begin{equation} \label{eq:int_rate}
    \dot{N}_{\mathrm{int}} 
    = k_{\mathrm{core}}  \, (m_1+m_2)  \,  m_3 \, a \,  \sqrt{m_1+m_2+m_3},
\end{equation}
where the scaling factor $k_{\mathrm{core}}$ depends on the local core properties, which we assume to be the same for all BBHs. 
We sample the original BBH that is most likely to interact based on its rate, and then apply the same procedure to determine whether the most interacting BBH is original or dynamical. 

We evolve the interacting BBH through a series of dynamical encounters until it is ejected by dynamical recoil or merges for GW emission. Then, we generate a new BBH. In the cluster core, BBHs can undergo binary-single (BBH-BH) and binary-binary (BBH-BBH) interactions, based on their relative rate (\citealp{marinpina2025}, see also \citealp{fronimos2026}):
\begin{equation} \label{eq:rate_ratio}
    \frac{\Gamma_{\mathrm{bb}}}{\Gamma_{\mathrm{bs}}} = \left(\frac{N_{\mathrm{BH}}}{10^2} \right)^{-1/3},
\end{equation}
where $\Gamma_{\mathrm{bs}}$ ($\Gamma_{\mathrm{bb}}$) is the rate of BBH-BH (BBH-BBH) encounters, and $N_{\mathrm{BH}}$ is the number of BHs in the core. In the following, we summarize our modeling for BBH-(B)BH interactions.


\subsubsection{BBH-BH interactions} 
In a BBH-BH interaction, we sample the mass of the BH interloper from Eq.~\ref{eq:int_rate}, by setting $m_1, \, m_2, \, a$ as the parameters of the interacting BBH. 
We model the resonant phase of the BBH-BH interaction as a sequence of intermediate states \citep{samsing2014,samsing2018,randoforastier2025}. At each state, we sample a new eccentricity from a thermal distribution and calculate the energy loss from the passage at pericenter \citep{hansen1972,samsing2014}. If the energy loss exceeds the BBH binding energy, the BBH merges for GW capture. 

If the BBH does not merge during the resonant phase, the BBH-BH interaction leads to either a fly-by or an exchange. We select the proper scattering outcome using the relative cross-sections by \cite{sigurdsson1995}, which span across a wide range of mass ratios and velocity regimes. We then evaluate the variation of the BBH semi-major axis and binding energy. For fly-bys, we consider the hardening prescription by \citet{randoforastier2025} that also accounts for post-Newtonian corrections. For exchanges, we sample the new semi-major axes following \citet{sigurdsson1995}. Finally, we generate the new BBH eccentricity from a thermal distribution and the new spin directions isotropically. In Appendix \ref{app:bbh_bh_int}, we report all the details and analytic formulas in our BBH-BH interaction model.



\begin{figure*}
    \centering
    \includegraphics[width=\hsize]{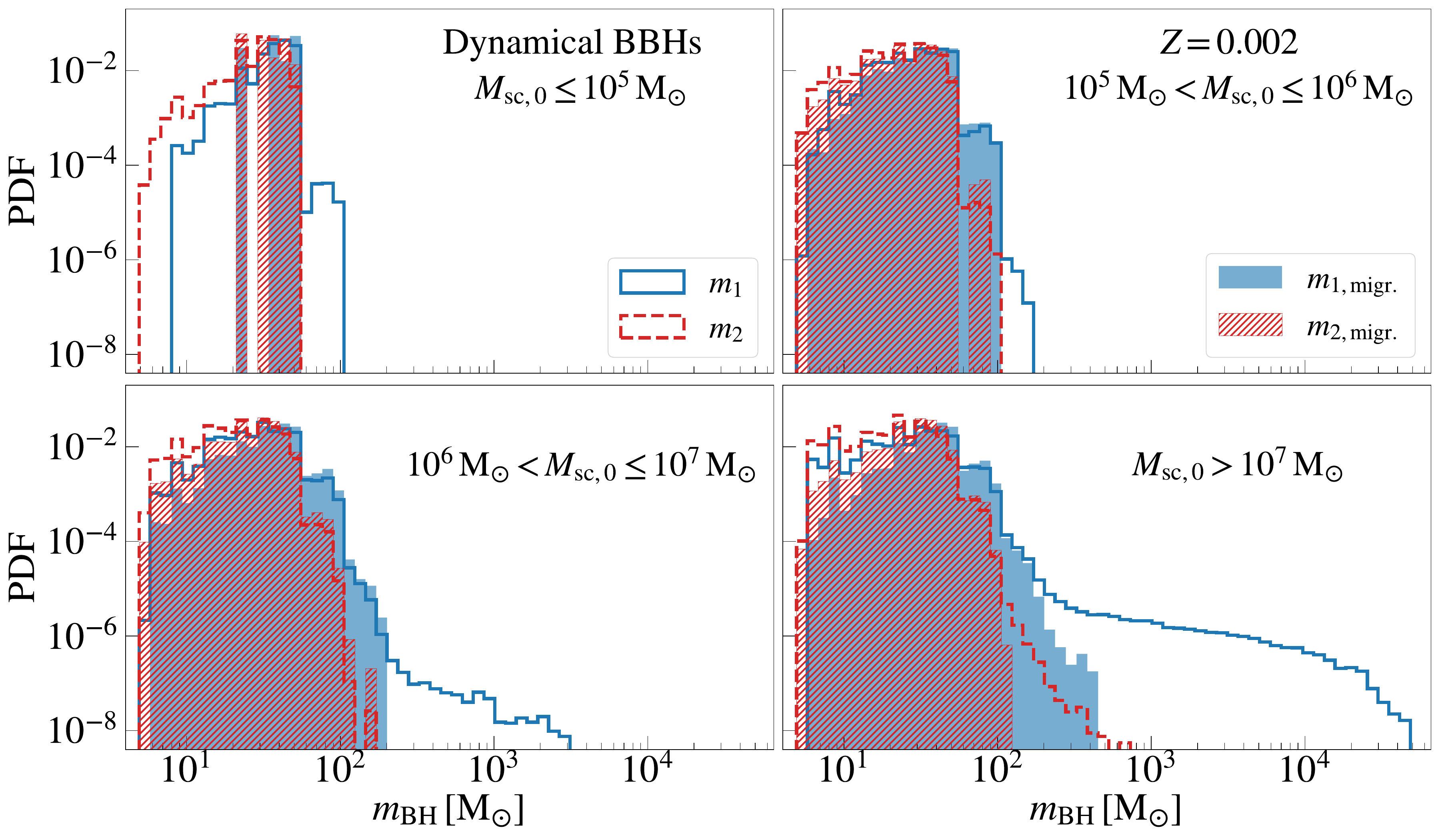}
    \caption{Primary (blue, solid) and secondary (red, dashed) mass distributions of dynamical BBH mergers. We show the distributions for all the star clusters (step) and for only those that migrate to the galactic center (filled, hatched). Different panels show star clusters with different masses. 
    }
    \label{fig:BBH_dynamical_mergers_hist}
\end{figure*}

\subsubsection{BBH-BBH interactions} \label{sec:bbh_bbh_int}
In a BBH-BBH interaction, we first generate a dynamical BBH interloper candidate. We draw the BBH component masses following the same procedure as Sect.~\ref{sec:dyn_orig_bbh}. We generate the interloper semi-major axis from a distribution $p(a_{\mathrm{max}} / a_{\mathrm{min}}) \sim (a_{\mathrm{max}} / a_{\mathrm{min}})^{-1}$, where $a_{\mathrm{max}}$ ($a_{\mathrm{min}}$) is the larger (smaller) semi-major axis in the interaction \citep{marinpina2025}.  
Since the BBH responsible for energy generation likely has already undergone dynamical hardening, we assume that the BBH interloper has always the larger semi-major axis. Then, we draw an original BBH interloper candidate, and determine if the BBH interloper is dynamical or original, in the same way as Sect.~\ref{sec:dyn_orig_bbh}. 

We evaluate the probability that a BBH undergoes a GW capture during the resonant phase using the fitting formula by \cite{marinpina2025}. If the BBHs do not merge, we sample the outcome of the BBH-BBH interaction based on the cross-sections from BBH-BBH scattering experiments by \cite{antognini2016}, in the proper semi-major axis ratio and velocity regime. The possible outcomes are fly-by, exchange, ionization, or triple formation. 

For each outcome, we calculate the dynamical BBH hardening as described in Appendix \ref{app:bbh_bbh_int}, we sample the new BBH eccentricity from a thermal distribution and we draw the new spin directions isotropically over a sphere. When a triple system forms, we assess if it merges by von Zeipel-Lidov-Kozai (ZLK, \citealp{vonzeipel1910,lidov1962,kozai1962}) oscillations before the next interaction. 
We report all the details on our BBH-BBH interaction model in Appendix \ref{app:bbh_bbh_int}.

\subsubsection{Hierarchical mergers} \label{sec:bbh_hardening}
After each interaction, we evaluate the dynamical recoil on the BBH and the interloper (B)BH. If it is larger than the star cluster escape velocity, we consider the (B)BH as ejected and estimate its GW merger timescale \citep{peters1964}. If the latter is shorter than the Hubble time, we consider the BBH as merged.  
If the BBH is retained, we integrate the hardening due to GW emission to estimate if it merges before the next dynamical interaction. At each time, we also evaluate if any original BBHs in the star cluster have merged due to GW hardening. 

When a BBH merges, we calculate the remnant mass, spin, and the relativistic recoil using fits from numerical relativity models, namely \texttt{NRSURur7dq4Remnant} \citep{varma2019} and \texttt{NRSur7dq4EmriRemnant} \citep{boschini2023} for $q \leq 0.2$\footnote{Since \texttt{NRSur7dq4EmriRemnant} does not include prescriptions for GW-induced kicks, we use the fitting formulas by \cite{lousto2012}.}. If the BBH merger occurs within the cluster and the relativistic recoil is smaller than the cluster escape velocity, the BH remnant is retained. In this case, we assume that the BH remnant is sent to the cluster outskirts and sinks back after a dynamical friction time \citep{antonini2019}. After that, the BH remnant can undergo dynamical interactions with the BH core population.

\subsection{Star cluster evolution} \label{sec:sc_evolution}
We model cluster evolution due to stellar mass-loss, two-body relaxation and tidal stripping using \textsc{clusterBH} \citep{antonini2020}. This model for star cluster evolution follows the two-component formalism by \cite{breen2013a}. The cluster consists of a stellar population with average mass $\bar{m}_{\mathrm{\star}}=0.6 \, \msun$ and total mass $M_{\mathrm{\star}}$, and a population of massive objects, namely BHs, with total (average) mass $M_{\mathrm{BH}}$ ($\bar{m}_{\mathrm{BH}}$). The total cluster mass is $M_{\rm{sc}} = M_{\mathrm{\star}}+M_{\mathrm{BH}}$. 
Here, we combine \textsc{clusterBH} with our model for BH dynamical interactions and mergers introduced in Sect. \ref{sec:bbh_core_init}. Also, we implement stellar mass loss due to tidal shocks, 3D orbit integration within the galactic potential and dynamical friction from the stellar, gas and dark matter component in the galaxy.

The total stellar mass decreases as a power law due to stellar mass loss by stellar winds and supernova explosions \cite{antonini2020}. We calibrate the cluster expansion to reproduce the evolution of the \textsc{Dragon-II} $N-$body simulations \citep{arcasedda2023d1}.  The total BH mass evolves as a result of BH dynamical ejections and mergers in the cluster core, as described in Sect. \ref{sec:bbh_core_init}. In our model, we update $M_{\mathrm{BH}}$ and $\bar{m}_{\mathrm{BH}}$ after each BBH–(B)BH encounter and BH merger. The time interval between successive interactions is evaluated under the assumption of balanced evolution between BBH hardening and global cluster expansion. The resulting expansion rate is derived assuming virial equilibrium, following \cite{antonini2020}.

In Appendix~\ref{app:internal_evolution}, we show that the adopted model successfully reproduces cluster evolution in $N$-body simulations. In the following, we focus on the cluster interaction with the host galaxy and the integration of cluster orbits within the galactic potential.

\subsubsection{Tidal stripping} \label{sec:tidal}
We model tidal stripping from the host galaxy using the prescription in \citet{gieles2023}, which also accounts for cluster expansion driven by BH heating:
\begin{equation}
\dot{M}_{\star,\rm tid} = -45 \, \mathrm{M_{\odot} \,Myr^{-1}}
\left( \frac{M_{\mathrm{sc}}}{M_{\mathrm{sc,i}}} \right)^{-1/3}
\left( \frac{M_{\mathrm{sc,i}}}{2\times10^{5} \,\mathrm{M_\odot}} \right)^{1/3}
\frac{\Omega_{\mathrm{tid}}}{0.32 \,\mathrm{Myr^{-1}}},
\end{equation}
where $M_{\mathrm{sc,i}} = 0.55 \, M_{\mathrm{sc,0}}$ is the cluster mass after most of stellar-evolution mass loss has occurred. We generalize this formalism to arbitrary galactic potentials by defining $\Omega_{\mathrm{tid}} = \lambda_1 - (\lambda_2 + \lambda_3)/2$, where $\lambda_i$ are the eigenvalues of the galactic tidal tensor \citep{pfeffer2018}, evaluated self-consistently from the potential.


\subsubsection{Tidal shocks} \label{sec:shocks}
When a star cluster goes through the galactic disk and bulge or the overdensities in the interstellar medium, the perturbation in the gravitational potential increases the energy of the bound stars, potentially leading to their escape \citep{spitzer1987}. 
The cluster response to such tidal shock can be modeled under the impulse approximation, in terms of the cluster internal properties and the local galactic tidal tensor \citep{gnedin1997,gnedin1999a,gieles2006, prieto2008,kruijssen2011,gieles2016}. The resulting mass loss is \citep{reinacampos2022}:
\begin{equation} \label{eq:tidal_shocks}
\begin{split}   
    \dot{M}_{\mathrm{\star,sh}} = - 27.1 \, \mathrm{\msun \, Myr^{-1}} \left( \frac{r_{\mathrm{h}}}{4 \, \mathrm{pc}} \right)^{3}  \left[ \sum_{ij} \left|\frac{\int T_{\mathrm{ij}}\mathrm{d}t}{10^2 \, \mathrm{Gyr^{-1}}}\right| A_{\mathrm{w,ij}} \right]  \\ \cdot \left( \frac{\Delta t_{\mathrm{sh}}}{10 \, \mathrm{Myr}} \right)^{-1},
\end{split}
\end{equation}
where $\Delta t_{\mathrm{sh}}$ is the time interval between two shocks and $A_{\mathrm{w,ij}}$ is the Weinberg adiabatic correction \citep{weinberg1994}, which accounts for the absorption of injected energy through the adiabatic expansion of the cluster:
\begin{equation}
    A_{\mathrm{w,ij}} = \left( 1 + \eta_{\mathrm{A}} \frac{G M_{\mathrm{sc}}}{r^3_{\mathrm{h}}} \tau^2_{\mathrm{ij}} \right)^{-3/2}.
\end{equation}
Here, $\eta_{\mathrm{A}}=0.24$ is a constant and $\tau_{\mathrm{ij}}$ is the shock duration for the corresponding component of the tidal tensor \citep{gnedin1997,gnedin1999b}. 
We perform the integral $\int T_{\mathrm{ij}} \, \mathrm{d}t$ over the shock duration for each tensor component, between minima with a sufficient contrast ($<0.88$) with respect to the bounded maximum, following \cite{pfeffer2018}. 

After a tidal shock, the half-mass radius evolves as:
\begin{equation}
    \dot{r}_{\mathrm{h,sh}} = \left( 2 - \frac{1}{f_{\mathrm{sh}}} \right) \frac{r_{\mathrm{h}}}{M_{\mathrm{sc}}} \dot{M}_{\mathrm{\star,sh}},
\end{equation}
where $f_{\mathrm{sh}}= |\mathrm{d} \ln{M_{\mathrm{sh}}}|/|\mathrm{d} \ln{E_{\mathrm{sh}}}|$ is the fraction of the relative energy change that is converted into the star cluster mass variation. We calculate $f_{\mathrm{sh}}$ self-consistently from \cite{gieles2016}, following \cite{reinacampos2022}.

\subsection{Star cluster migration} \label{sec:sc_in_galaxies}
We implement the galactic potential and integrate the cluster orbit using \texttt{galpy} \citep{bovy2015, webb2019}. We combine the orbital integration with an adaptive time scheme to progressively update the cluster global properties (see Sect. \ref{sec:sc_evolution}) as it evolves within the host galaxy. In the presence of tidal shocks, we adjust the time step to have the required resolution to capture the tidal tensor variation at the disk passage. In this way, we combine the cluster progressive dissolution with its orbital evolution in the galactic tidal field. Finally, we include dynamical friction from the stellar, gaseous, and dark matter components of the galaxy.

\begin{figure}
    \centering
    \includegraphics[width=\hsize]{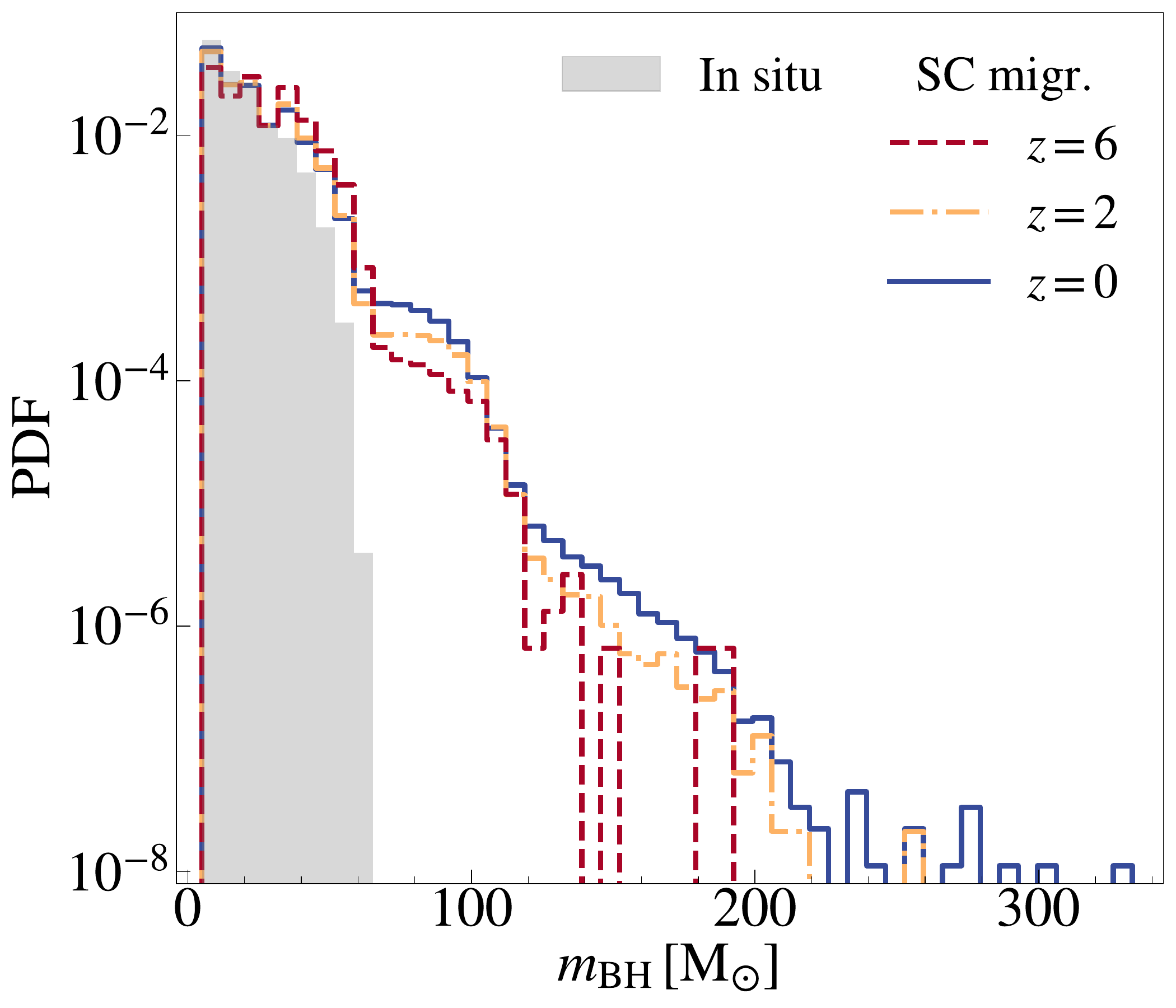}
    \caption{Mass distribution for BHs that populate galactic nuclei from star cluster migration within a certain redshift (from left to right): $z=6$ (red, dashed), $z=2$ (yellow, dashdot), and $z=0$ (blue, solid). The grey-filled histogram represents the BH initial mass distribution from in situ formation in galactic centers.}
    \label{fig:BHs_galnuc}
\end{figure}

\subsubsection{Dynamical friction} 
As star clusters evolve in the galactic tidal field, encounters with lighter components remove energy and momentum from the orbit \citep{chandrasekhar1943}, possibly driving their migration to the galactic center. We implement dynamical friction from the collisionless spherical components of the galaxy, namely stellar and dark matter haloes, as \citep{chandrasekhar1943,petts2016}:
\begin{equation} \label{eq:df}
\begin{split}    
    \textbf{F}_{\mathrm{df,\star}}(\textbf{x},\textbf{v}) = -2 \, \pi \, G^2 M^2_{\mathrm{sc}} \, \rho_{\mathrm{\star}}(\textbf{x})  \log{(1+\Lambda^2)} \\  \cdot \left[ \mathrm{erf(X)  - \frac{2X}{\sqrt{\pi}}} \exp(-X^2)  \right] \frac{\textbf{v}}{\textbf{|v|}^3}, 
\end{split}
\end{equation}
where $\textbf{x}$ and $\textbf{v}$ are the position and relative velocity of the cluster in the galaxy, $X=|\textbf{v}|/[\sqrt{2} \sigma(r)]$ and $\sigma_{\mathrm{gal}}(r)$ is the velocity dispersion, computed from the galactic potential self-consistently. We calculate the Coulomb logarithm following \cite{petts2016}:
\begin{equation}
    \Lambda(r) = \frac{r/\gamma_{\mathrm{g}}}{\max(r_{\mathrm{h}}, \, G M_{\mathrm{sc}}/|\textbf{v}|^2)},    
\end{equation}
where $\gamma_{\mathrm{g}}= |\mathrm{d}{\log{\rho_{\mathrm{\star}}}} / \mathrm{d}{\log{r}} |$ is the slope of the density profile. 

For disk stars, we model the effect of the axisymmetric rotating potential following \cite{bonetti2020}. This model is based on Eq.~(\ref{eq:df}), with the perturber velocity replaced by the relative velocity between the perturber and the disk, $|\textbf{v}|\rightarrow|\textbf{v}-\textbf{v}_{c}\mathrm{(R_{\mathrm{g}})}|$. Here,
$\textbf{v}_{\mathrm{c}}\mathrm{(R_{\mathrm{g}})}$ is the circular velocity of the disk stars at  $R_{\mathrm{g}}$, calculated self-consistently from the disk potential. 


Dynamical friction by the collisional gaseous medium acts differently than the collisionless stellar distribution. The gaseous drag is enhanced for supersonic motions, while it is less efficient in the subsonic regime. We model gas friction as \citep{ostriker1999}:
\begin{equation} \label{eq:df_g}
    \textbf{F}_{\mathrm{df,gas}}(\textbf{x},\textbf{v}) = -2 \, \pi \, G^2 M^2_{\mathrm{sc}} \, \rho_{\mathrm{gas}}(\textbf{x}) \, I(\mathcal{M}) \, \frac{\textbf{v}}{\textbf{|v|}^3}, 
\end{equation}
where $I$ is a function of the Mach number $\mathcal{M}=|\mathbf{v}|/c_{\mathrm{s}}$, and $c_{\mathrm{s}}$ 
is the sound speed of the gaseous medium:  
\begin{multline}
I_{\mathrm{subsonic}}(\mathcal{M}) = \frac{1}{2} \log{\left(\frac{1+\mathcal{M}}{1-\mathcal{M}}\right)} - \mathcal{M}, \hfill \\
I_{\mathrm{supersonic}}(\mathcal{M}) = \frac{1}{2} \log{\left(1-\frac{1}{\mathcal{M}^2}\right)} +  \log{(\Lambda)}.  \hfill
\end{multline}

\begin{figure}
    \centering
    \includegraphics[width=\hsize]{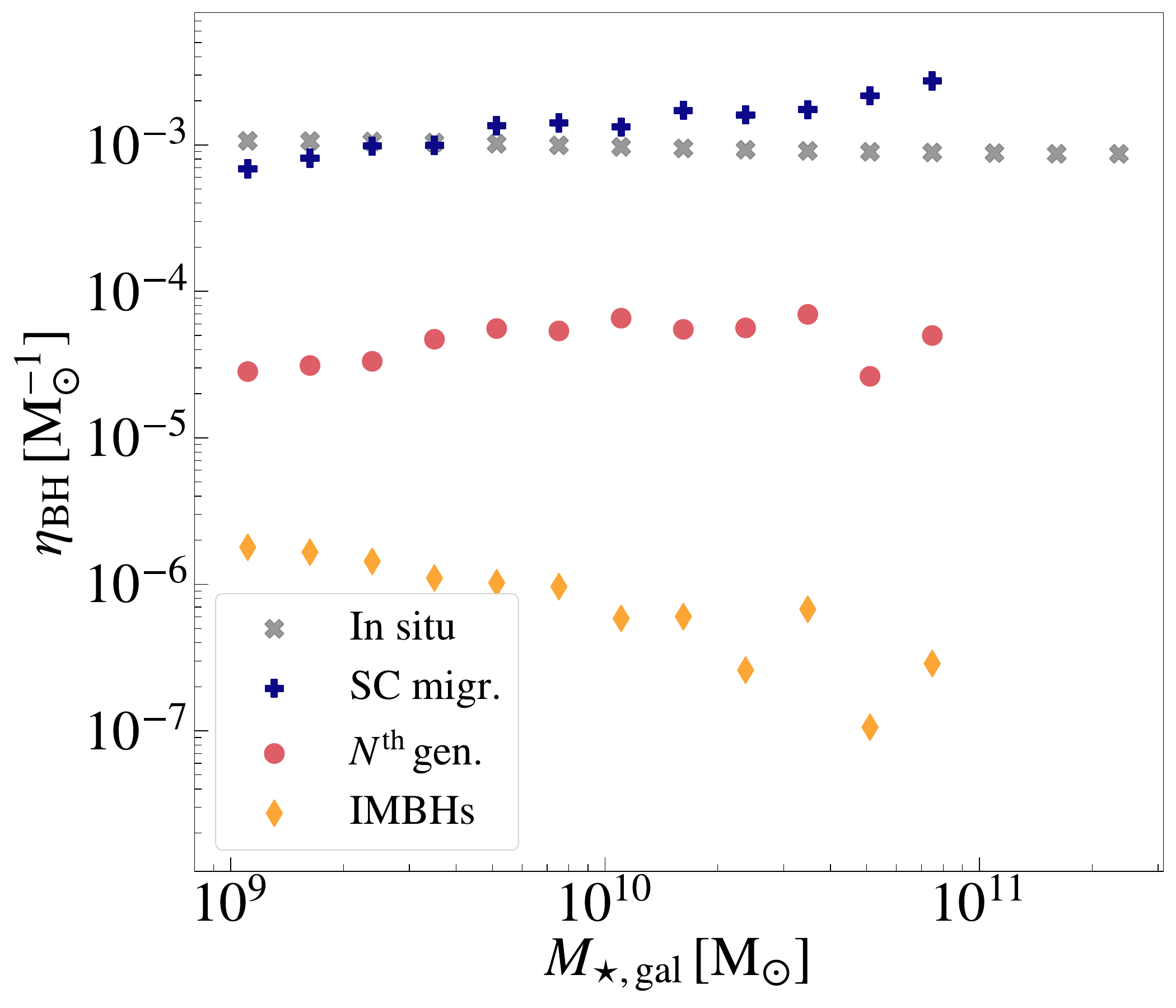}
    \caption{BH formation efficiency in galactic nuclei from different formation channels. Different colors represent different populations in the star cluster migration scenario: all BHs (blue, plus), $N^{\mathrm{th}}$-generation BHs (red, circle), IMBHs (yellow, diamond). The grey crosses represent the formation efficiency in the in situ formation scenario.}
    \label{fig:BHs_efficiency}
\end{figure}

\section{Results} \label{sec:results}

We evolve the star clusters until they dissolve, reach the galactic center, or until the Hubble time. If all the BHs are ejected from the cluster core, we continue the orbital integration to assess if the cluster can migrate to the galactic center. Since our focus is on the initial BH populations in galactic nuclei, we do not evolve star clusters and BHs after they reach the galactic center.

\subsection{Star cluster migration to the galactic center}
Figure \ref{fig:mcl_rg_scatter} shows the  distribution of initial mass and galactocentric distance, for disk and halo clusters. To enable a direct comparison across galaxies, we normalize these quantities by the host galaxy stellar mass ($M_{\mathrm{\star,gal}}$) and length scale ($R_{\mathrm{\star,gal}}$). 
We find a significant population of migrating star clusters at the high-mass end of the distribution. The ratio $M_{\mathrm{sc,0}}/M_{\mathrm{\star,gal}}$ primarily determines whether a star cluster reaches the galactic center 
(see Eqs.~\ref{eq:df} and \ref{eq:df_g}). The initial galactocentric distance has a fundamental impact on the migration timescale.

Star clusters can populate galactic nuclei as early as $z \sim 8$. Cluster migration through dynamical friction is possible only for the highest mass ratios, with $M_{\mathrm{sc,0}} \, / \, M_{\mathrm{\star, gal}} > 10^{-4}$.
The average migration timescale is $\sim 1 \, \mathrm{Gyr}$, with a distribution ranging from $10^2 \, \mathrm{Myr}$ for $R_{\mathrm{g,0}} / R_{\mathrm{\star, gal}} \sim 0.1$ up to the Hubble time for $R_{\mathrm{g,0}} / R_{\mathrm{\star, gal}} > 1 $. 
For migrating clusters, we find a relation: 
\begin{equation}
    \log{(M_{\mathrm{sc,0}} /M_{\mathrm{\star, gal}})} \sim \log{(R_{\mathrm{g,0}} /R_{\mathrm{\star, gal}})}^{0.95},
\end{equation}
with a residual dispersion of $0.25$.

Disk clusters also exhibit a second population of migrating systems, with $M_{\mathrm{sc,0}}/M_{\mathrm{\star,gal}} < 10^{-4}$. These clusters form close to the galactic center and sink rapidly after their formation ($t_{\mathrm{DF}} \ll 100 \, \mathrm{Myr}$). Given their small escape velocities, they retain a low BH fraction. As a result, they drag to the galactic center only a limited number of BHs, born from stellar collapse.


\subsection{Migration efficiency} \label{sec:migr_efficiency}
Figure \ref{fig:m0_mf_hist} shows the initial and final cluster mass distribution, for different galactic masses. More than $90\%$ of clusters with $M_{\mathrm{sc, 0}} \lesssim 10^{6} \, \msun$ are disrupted. Tidal shocks significantly contribute to their dissolution, and become efficient once clusters expand to $r_{\mathrm{h}} \gtrsim 4 \, \mathrm{pc}$ (see Eq.~\ref{eq:tidal_shocks}). For $M_{\mathrm{sc,0}} \gtrsim 10^6 \, \msun$, about $20\%$ of clusters either survive or reach the galactic center.

In galaxies with $M_{\mathrm{\star, gal}} \leq 10^{10} \, \msun$, the fraction of infalling clusters approaches unity when $M_{\mathrm{sc, 0}} > 5 \times 10^{6} \, \msun$, independently of the initial galactocentric distance. In this regime, rapid dynamical friction allows clusters to retain up to $40\%$ of their initial mass, while their compact initial size limits the impact of tidal shocks on their dissolution.
For $M_{\mathrm{\star, gal}} > 10^{10} \, \msun$, the most massive star clusters cannot reach the galactic center. Despite their large masses, these clusters exhibit moderate mass ratios ($M_{\mathrm{sc,0}} \,/ \, M_{\mathrm{\star, gal}} \sim 10^{-4}$), resulting in dynamical friction timescales that exceed the Hubble time.


Figure \ref{fig:migr_efficiency} shows the fraction of migrating clusters, $f_{\mathrm{migr}}$, as a function of the cluster and galactic mass. 
For a given galactic mass, $f_{\mathrm{migr}}$ is close to unity for the most massive star clusters that form in a galaxy, and decreases down to $10^{-4}$ for  $M_{\mathrm{sc,0}} \sim 10^5 \, \msun$. Lower-mass clusters can reach the galactic nucleus only if they form at very small galactocentric distances. In this regime, the migration efficiency is independent of galactic mass and reflects the probability of forming near the center, rather than the effect of dynamical friction. Independently of the star cluster mass,
the migration efficiency drops to zero for $M_{\mathrm{\star, gal}} > 10^{11} \, \msun$.


\subsection{BBH mergers} \label{sec:bbh_mergers}

We classify BBH mergers into original and dynamical BBHs, based on their formation channel. Original BBHs form from stellar binaries that are bound at birth and do not undergo dynamical interactions. Dynamical BBHs form through three-body interactions or exchange, and subsequently harden through dynamical encounters. 
We show our distributions of original BBH mergers in Appendix \ref{sec:original_mergers}. In the following, we focus on dynamical and hierarchical BBH mergers.



Figure \ref{fig:BBH_dynamical_mergers_hist} shows the mass distribution of dynamical BBH mergers in clusters with different masses, at $Z=0.002$. The mass distribution shows a main peak at $\sim 50 \, \msun$, which corresponds to the lower edge of the pair-instability mass gap \citep{iorio2023}. As explained in Sect. \ref{sec:dyn_orig_bbh}, the adopted pairing criterion, combined with exchanges, reproduces the tendency of dynamical encounters to couple and harden the most massive BHs. This results in a peak at the upper end of the first-generation mass distribution (e.g., see \citealp{antonini2023,torniamenti2024}). Hierarchical mergers overcome this upper limit, producing a high-mass tail that depends strongly on cluster mass.


Star clusters with $M_{\mathrm{sc,0}} < 10^6 \, \msun$ produce hierarchical chains up to the second generation, resulting in a secondary peak at around $100 \, \msun$. Higher-order generations are suppressed by the GW-induced recoils, which eject BH remnants from the cluster \citep{gerosa2021}. For $M_{\mathrm{sc,0}} > 10^6 \, \msun$, hierarchical BH mergers progressively populate the intermediate-mass range above $10^3 \, \msun$. In the most massive clusters ($M_{\mathrm{sc,0}} > 10^7 \, \msun$), the large masses and densities ($\rho_{\mathrm{h}} \gtrsim 10^{6}-10^{7} \, \msun \, \mathrm{pc}^{-3}$) result in escape velocities larger than $500 \, \mathrm{km \, s^{-1}}$. This significantly enhances the retention of BH remnants and enables the formation of hierarchical chains with $\sim 10^2-10^3$ generations. As a result, hierarchical mergers can drive the formation of BHs exceeding $10^4 \, \msun$ (see also \citealp{chattopadhyay2023,kritos2023}).

BBH mergers with $m_{\mathrm{BH}} > 10^3 \, \msun$ occur in the most massive clusters in galaxies with $M_{\rm{\star,gal}} > 10^{10} \, \msun$, which do not spiral into the galactic center (see Fig. \ref{fig:m0_mf_hist}). The growth of such massive BHs occurs over timescales of a few Gyr, longer than the typical migration timescale. As a consequence, migrating star clusters only exhibit a BBH merger distribution that extends to $400 \, \msun$. In Sect. \ref{sec:sc_birth_radii}, we discuss how this result depends on our assumptions on the initial cluster radius.

\begin{figure}
    \centering
    \includegraphics[width=\hsize]{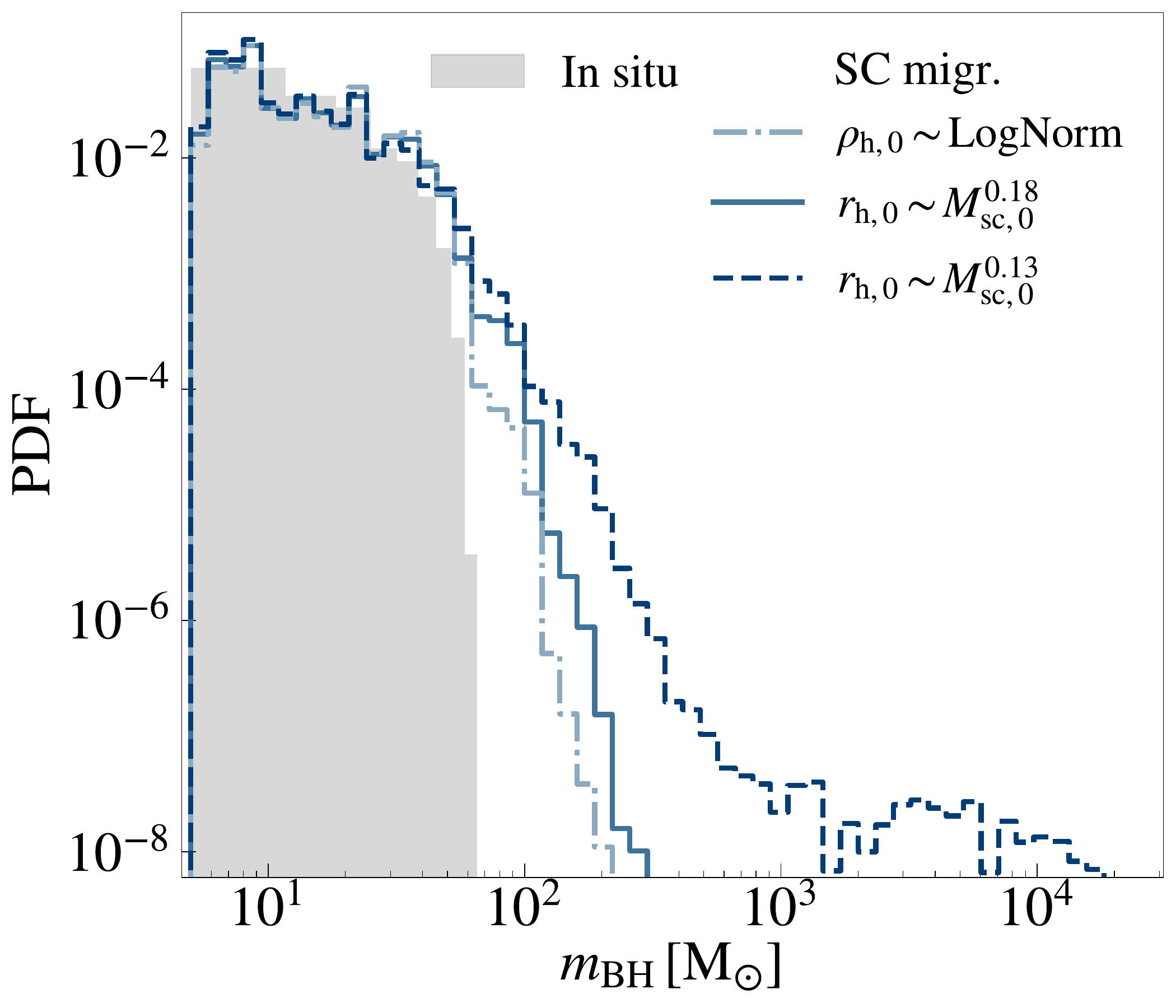}
    \caption{Mass distribution for BHs that populate galactic nuclei from star cluster migration, for models initialized with different birth radii: \cite{rantala2024} (blue, solid), \cite{markskroupa2012} (dark blue, dashed), and a Log-normal distribution for the half-mass density (\citealp{mapelli2021}, light blue, dash-dot). The grey-filled histogram represents the BH initial mass distribution from in situ formation.
    }
    \label{fig:BHs_rhm}
\end{figure}

\subsection{BH populations in the galactic center}

Figure \ref{fig:BHs_galnuc} shows the BH mass distribution in galactic nuclei from migrating star clusters. As a reference, we show the mass distribution for BHs that form in situ. Specifically, we consider the metallicity distribution in the central annuli (within $30 \, \mathrm{pc}$) from the \textit{L-Galaxies} catalogs for galaxies where cluster migration occurs to generate the population of first-generation BHs that form in situ. Our models indicate that star cluster migration can seed galactic nuclei with IMBHs as early as redshift $z \sim 6$. 

In the in-situ formation scenario, the upper limit of the BH mass distribution is $60 \, \msun$. The BH mass distribution from cluster migration, instead, extends up to the intermediate-mass regime, thanks to the retention of $N^{\mathrm{th}}$-generation BHs. The maximum BH mass increases with decreasing redshift, as clusters with longer migration times populate galactic nuclei. At $z=6$, BH mergers in migrating clusters fill the mass range between $50-100 \, \msun$ with second-generation BHs. 
At $z=0$, the upper limit of the BH mass distribution exceeds $m_{\mathrm{BH}}=300 \, \msun$, a factor of five higher than the maximum BH mass from in-situ formation. IMBHs up to the tenth generation account for $0.4 \%$ of the total BH mass, while $N^{\mathrm{th}}$-generation BHs represent $\sim 5\%$.



\subsection{BH formation efficiency in galactic nuclei}

{
Figure \ref{fig:BHs_efficiency} shows the BH formation efficiency ($\eta_{\mathrm{BH}}$) in NSCs, for different formation channels. Overall, in situ formation and star cluster migration result in comparable efficiencies, with mildly different dependencies on the galactic mass. 
In the cluster migration scenario, $\eta_{\mathrm{BH}}$ increases with the galactic stellar mass, because longer dynamical friction timescales lead to enhanced stellar mass loss during the inspiral. In galaxies with $M_{\mathrm{\star, gal}} > 10^{11} \, \msun$, the formation efficiency vanishes, because the migration timescales exceed the Hubble time (see, e.g., Fig. \ref{fig:migr_efficiency}). 
}



{
The IMBH formation efficiency is $2 \times 10^{-6} \, \msun^{-1}$ for $M_{\mathrm{\star, gal}} < 10^{10} \, \msun$, where the nucleation peak resides {\citep{sanchezjanssen2019,neumayer2020,hoyer2021}}. In more massive galaxies, higher metallicities lead to lower BH masses. As a consequence, IMBH migration efficiency progressively decreases, despite the presence of a non-negligible population of $N^{\mathrm{th}}$-generation BHs. The BH migration efficiency of $N^{\mathrm{th}}$-generation BHs is approximately $3 \times  10^{-5} \, \msun^{-1}$, with a weak dependence on the galaxy mass. 
}

\section{Impact of star cluster birth radii and BH spins} \label{sec:discussion}
Hierarchical BH mergers involve a number of unconstrained parameters, related to both BH physics and star cluster formation. In the following, we encompass these uncertainties by exploring two key factors that affect IMBH formation and retention in hierarchical mergers: the initial cluster size and BH spins.

\subsection{Star cluster birth radii} \label{sec:sc_birth_radii}
The birth radii of star clusters are still highly unconstrained, but they have a fundamental impact on their dynamical evolution. Smaller initial radii imply shorter dynamical timescales and larger initial escape velocities, thus enhancing the rate of hierarchical mergers. In this work, we initialize our star cluster models following \cite{rantala2024} (see Eq. \ref{eq:rantala24}), who parameterize the initial mass-radius relation to reproduce the small observed birth size of embedded clusters, and capture the existence of massive parsec-size clusters in the early Universe \citep{adamo2024}.

We quantify the impact of our assumptions by considering two alternative prescriptions. First, we consider a \cite{markskroupa2012} relation, which is conceived to describe star clusters with $M_{\mathrm{sc}}\sim 10^4-10^6 \, \msun$ in their embedded phase:
\begin{equation} \label{eq:mandk2012}
    \frac{r_{\mathrm{h,0}}}{\mathrm{pc}} = 0.10^{+0.07}_{-0.04} \times \left( \frac{M_{\mathrm{sc,0}}}{\msun} \right)^{0.13 \pm 0.04}.
\end{equation}
By extending this relation to the entire mass range of our sample, we can obtain massive ($M_{\mathrm{sc,0}} \gtrsim 10^6 \, \msun$) and very compact ($r_{\mathrm{h}} \lesssim 0.5 \, \mathrm{pc}$) star clusters.
We also consider an initial half-mass density log-normal distribution with mean $\langle{}\log_{10}{\rho{}/({\rm M}_\odot\,{}{\rm pc}^{-3})}\rangle{}=4.5$, with a standard deviation $\sigma_\rho=0.4$ \citep{mapelli2022}. In this case, we do not assume any dependence between the initial star cluster mass and half-mass radius. This results in the most massive clusters having large initial radii, with $r_{\mathrm{h}} \gtrsim 2 \, \mathrm{pc}$. 
With these assumptions, we can encompass the uncertainties related to the birth radii of massive clusters ($M_{\mathrm{sc,0}} \gtrsim 10^6 \, \msun$), spanning from sub-parsec scales to values $> 5 \, \mathrm{pc}$.

Figure \ref{fig:BHs_rhm} shows the BH populations in galactic nuclei in models with different $r_{\mathrm{h,0}}$ distributions. 
Star clusters initialized with a log-normal density distribution present the lowest escape velocities. Also, they are more easily disrupted by tidal shocks, given their large birth size. This quenches the IMBH formation and migration, with an IMBH mass fraction in galactic nuclei that decreases to $0.1\%$ and a maximum BH mass $m_{\mathrm{BH}} \sim 280 \, \msun$.

Star clusters initialized with a \cite{markskroupa2012} relation
exhibit a significantly higher efficiency of IMBH formation and migration. Here, migrating clusters produce massive BH seeds above $10^4 \, \msun$, with IMBHs of up to $1000 \, \msun$ already present in galactic nuclei as early as $z \sim 6$. 
As shown in Fig. \ref{fig:BBH_dynamical_mergers_hist}, IMBHs exceeding $10^4 \, \msun$ can form in the most massive non-migrating clusters, even when initialized with larger birth radii (Eq. \ref{eq:rantala24}). 
If $r_{\mathrm{h,0}} \lesssim 0.5 \, \mathrm{pc}$, however, such massive IMBHs can form in migrating clusters, thanks to the enhanced escape velocities and shorter dynamical times. As a result, IMBHs above $10^4 \, \msun$ can form within their parent cluster and populate galactic nuclei.

\begin{figure}[ht]
    \centering
    \includegraphics[width=\hsize]{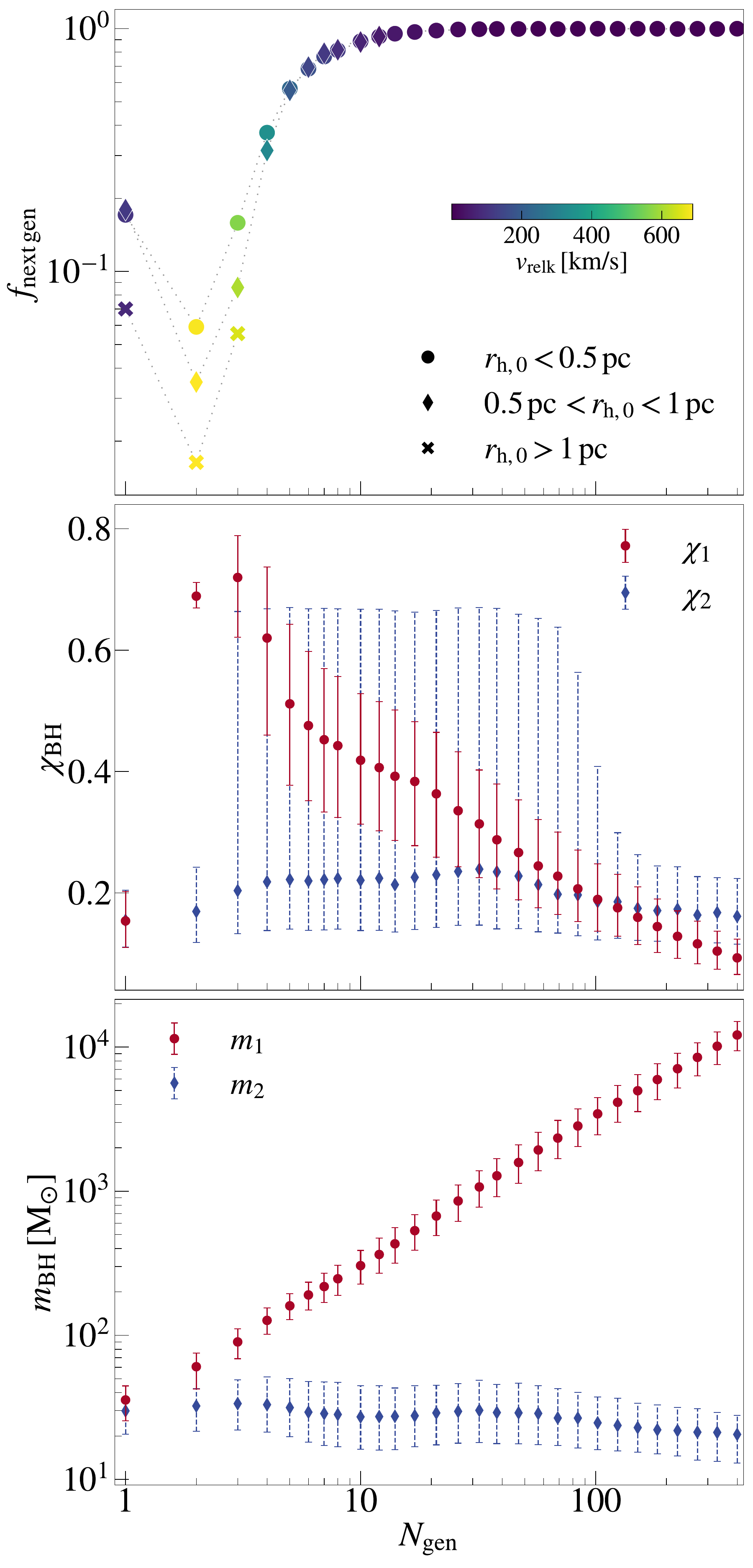}
    \caption{
    Hierarchical BH assembly in migrating clusters. Upper panel: fraction of BBH mergers that produce a merger at the successive generation ($f_{\mathrm{next \, gen}}$), for different initial radii. The color encodes the median GW-induced kick. Middle panel: median primary (red, circle) and secondary (blue, diamond) BH spin for each BH merger generation. Lower panel: median primary (red, circle) and secondary (blue, diamond) BH mass. The errorbars encompass the $25^{\mathrm{th}}$ and the $75^{\mathrm{th}}$ percentiles.}
    \label{fig:f_gen}
\end{figure}


\subsubsection{Hierarchical chains in migrating clusters}
In Figure \ref{fig:f_gen}, we investigate how the cluster birth radius affects the hierarchical BH assembly in migrating clusters. At each generation, we evaluate the fraction of hierarchical mergers that survive to the next generation ($f_{\mathrm{next \, gen}}$) and the median GW-induced recoil velocity ($v_{\mathrm{relk}}$), in clusters with different initial radii.

The early generations ($N_{\mathrm{gen}}=2-3$) represent the main bottleneck for the hierarchical BH assembly. 
At the first generation, BH remnants undergo relatively small GW-induced kicks, and can be retained in star clusters with $v_{\mathrm{esc}} \lesssim 100 \, \mathrm{km \, s^{-1}}$. At the second generation, the median recoil velocity exceeds $600 \, \mathrm{km \, s^{-1}}$ due to the sharp increase in the primary BH spin ($\chi_{\mathrm{BH}} \sim 0.7$). This decreases the retention fraction by at least a factor three.

After the second generation, the primary spin progressively decreases, due to repeated mergers with isotropically-oriented spins \citep{berti2008}. At the same time, the BBH merger mass ratio drops, since most interactions involve first-generation secondaries, as indicated by the nearly flat spin and mass trends. As a result, the GW-induced recoils become weaker and the BH remnant ejection increasingly inefficient. Combined with the shorter dynamical-friction times required for BHs to return to the cluster core, this speeds up the hierarchical assembly, driving the formation of IMBHs exceeding $10^4 \, \msun$.

If migrating clusters form at sub-parsec scales, the hierarchical BH assembly overcomes the first generations thanks to the large escape velocities ($v_{\mathrm{esc}} > 500 \, \mathrm{km \, s^{-1}}$). 
If $r_{\mathrm{h,0}} \lesssim 0.5 \, \mathrm{pc}$, the BH mass growth proceeds rapidly enough to populate galactic nuclei with such massive IMBHs. If $r_{\mathrm{h,0}}> 1 \, \mathrm{pc}$, the BH assembly stops at the third generation, even in the most massive clusters.

Constraining the initial size of star clusters is challenging, as the densest systems can expand by more than an order of magnitude within the very first Myr (see also \citealp{vergara2025}). Current \textit{JWST} observations of proto-star clusters \citep{adamo2024,claeyssens2026} suggest the existence of massive systems with parsec-scale sizes, supporting hierarchical mergers as a viable channel for seeding galactic nuclei with IMBHs.

\subsection{BH spins} \label{sec:disc_bhspins}
The relativistic recoil that BH remnants experience strongly depends on the BH spin magnitude. Specifically, higher spins lead to stronger relativistic kicks, possibly quanching the retention of BH remnants. In our fiducial model, we have considered a Maxwellian distribution with $\sigma_{\chi}=0.1$, which is reminiscent of the distribution inferred from GW detections. In the following, we explore the cases of first-generation BHs with spins very close to zero ($\sigma_{\chi}=0.01$) or relatively high ($\sigma_{\chi}=0.2$).

\begin{figure}
    \centering
    \includegraphics[width=\hsize]{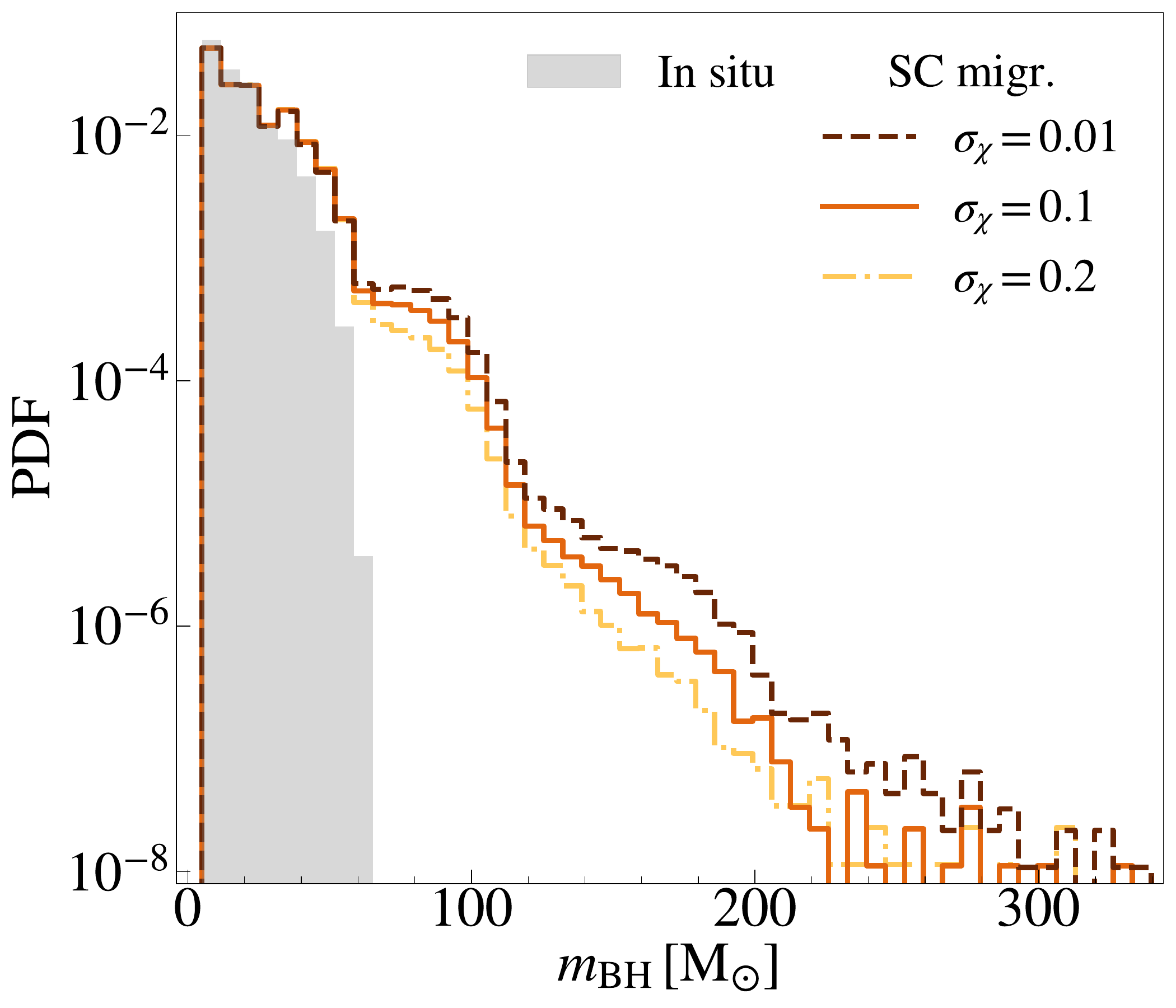}
    \caption{Mass distribution of BHs that migrate to galactic nuclei, for models initialized with Maxwellian BH spin distributions with $\sigma_{\chi}=0.2$ (yellow, solid), $\sigma_{\chi}=0.1$ (orange, dashdot), $\sigma_{\chi}=0.01$ (brown, dashed). The grey-filled histogram represents the initial mass distribution of BHs that form in situ in galactic centers.}
    \label{fig:BHs_spins}
\end{figure}

Figure \ref{fig:BHs_spins} shows how the BH populations in galactic nuclei depend on our assumptions on the initial spin distribution, for our fiducial model. Initial spin values close to zero increase the fraction of $N^{\mathrm{th}}$-generation BHs in galactic nuclei by a factor of three, {due to a more efficient remnant retention at the first generation}. Also, smaller (larger) BH spins increase (decrease) the IMBH fraction by a factor of 2 with respect to our fiducial case. 

Our assumptions on BH spins only have a minor impact on the maximum BH mass in galactic nuclei. In fact, second-generation BHs predominantly feature a narrow distribution around $\chi_{\mathrm{BH}} \sim 0.7$ (see Fig. \ref{fig:f_gen}). As a result, the hierarchical BH assembly at successive generations is less sensitive to the initial BH spins, and primarily depends on the cluster escape velocity.


\section{Discussion} \label{sec:caveats}

\subsection{Comparison with GW events} \label{sec:mbh_chibh}

We qualitatively compare our models with the primary components of GW190521 \citep{abbottGW190521}, GW231123 \citep{gw231123}, GW241011 and GW241110 \citep{gw241011}. These are among the most massive and/or highly spinning GW events, which suggest a possible hierarchical origin. 

\begin{figure}
    \centering
    \includegraphics[width=\hsize]{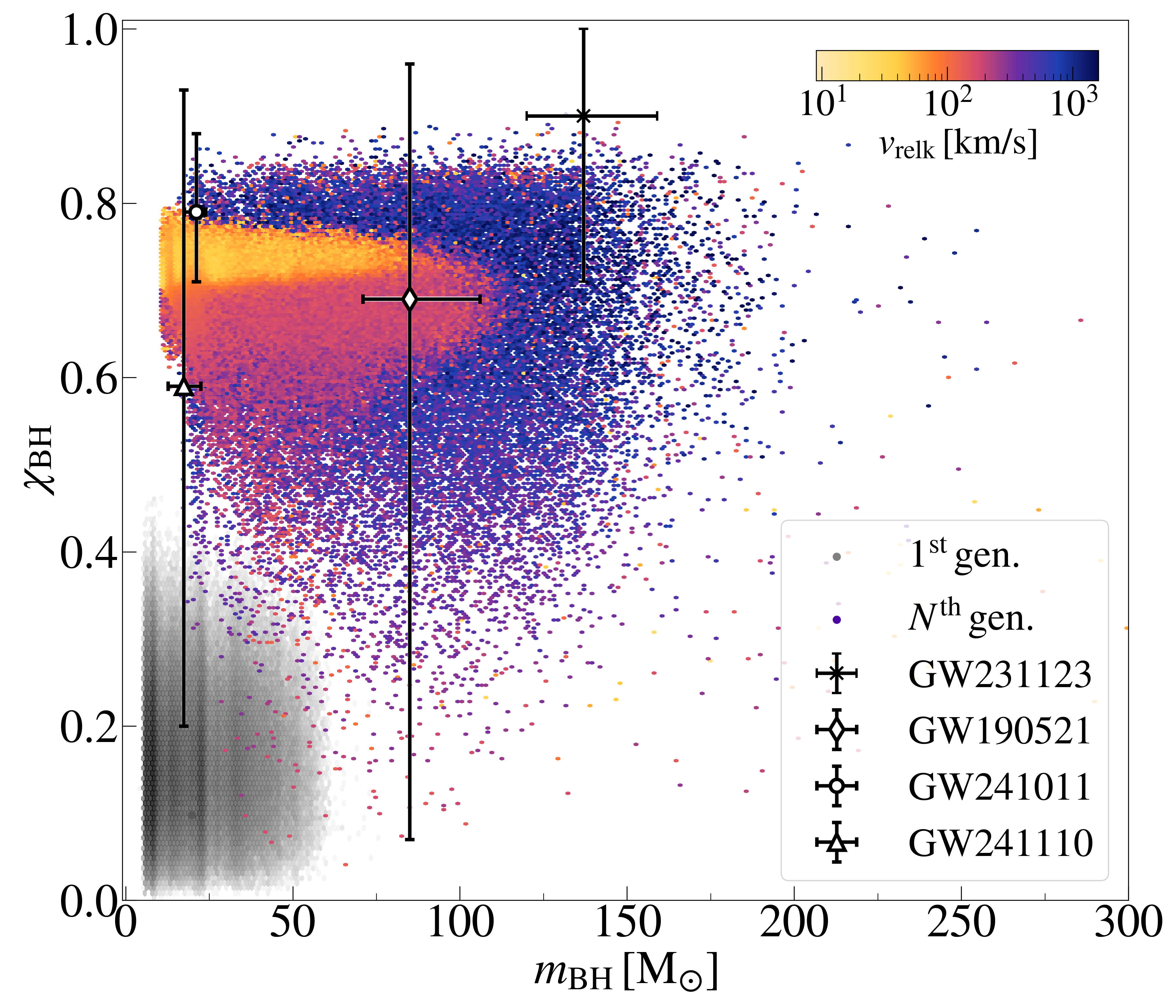}
    \caption{Mass ($m_{\mathrm{BH}}$) and spin ($\chi_{\mathrm{BH}}$) distributions for first- (grey) and $N^{\mathrm{th}}$- generation BHs. The color represents the relativistic kicks that BH remnants have experienced at formation. The error bars represent the primary mass and spin components of four gravitational-wave detections, namely GW231123 (cross), GW190521 (diamond), GW241011 (circle), and GW241110 (triangle).}
    \label{fig:ng_vk_scatter}
\end{figure}

Figure \ref{fig:ng_vk_scatter} shows the $m_{\mathrm{BH}}-\chi_{\mathrm{BH}}$ distribution for first- and $N^{\mathrm{th}}-$generation BHs in our models. For the latter, GW-induced recoil velocities span from $10 \, \mathrm{km \, s^{-1}}$ to $2000 \, \mathrm{km \, s^{-1}}$.  
The weakest kicks ($\lesssim 50 \, \mathrm{km/s}$) occur for remnants with $\chi_{\mathrm{BH}} \sim 0.7$, which originate from first-generation BBH mergers with initially low- and aligned spins (see \citealp{borchers2025}). The recoil velocities in $N^{\mathrm{th}}$-generation mergers can exceed $1000 \, \mathrm{km \, s^{-1}}$, due to the presence of at least one highly-spinning component.

Most of the GW events considered are broadly consistent with a hierarchical formation scenario in star clusters. Their primary masses and spins are well reproduced by BH remnants undergoing relatively small kicks ($\lesssim 100 \, \mathrm{km \, s^{-1}}$), and can therefore be retained within their parent cluster. GW231123 is consistent with a third-generation primary and a second-generation secondary \citep{passenger2025,angeloni2026b}. In our models, such GW231123-like events occur only in the most massive clusters, with $M_{\mathrm{sc,0}} > 2 \times 10^{6} \, \msun$.

\subsection{Possible limitations}


\subsubsection{Impact of galaxy mergers}

The initial galactocentric distance plays a fundamental role in determining the dynamical friction timescale.
For disk clusters, this number is randomly sampled from a uniform distribution within the annulus \citep[see][for details]{hoyer2025}. In combination with the high statistical number of star clusters, we believe that our results do not strongly depend on this assumption.

Halo clusters originate from the redistribution of clusters during galaxy mergers. In minor mergers ($q_{\mathrm{gal}} < 0.1$), this process typically moves them to larger galactocentric radii, depending on the distance and masses of the merging galaxies, as detailed in \citet{hoyer2025}. As a result, most accreted star clusters do not reach the galaxy center and therefore do not contribute to the central BH population. 
During minor mergers, \textit{L-Galaxies 2020} assumes that star clusters in the central galaxy remain unaffected. This assumption is supported by, for example, \citet{hilz2012a}, who show that the specific energy distribution of particles in the central galaxy does not significantly change.

In a major galaxy merger ($q_{\mathrm{gal}} \geq 0.1$), the disk of the central galaxy is destroyed and all star clusters populate the stellar halo.
For simplicity, \citet{hoyer2025} assume that clusters redistributed from the disk to the halo retain their original galactocentric distance during this process. This assumption may affect our results, as we expect changes in the radial action of star clusters during this phase. For example, \citet{del-lagos2018} show that the specific angular momentum of stars may change by up to $30 \%$, based on numerical results from the \textsc{Eagle} simulation.

Within the selected galaxy sample from the \textit{L-Galaxies 2020} simulation, we find that galaxy mergers are most important between redshifts $z=2$ and $z=0$, i.e.\ the presented results at $z=6$ are unaffected by any uncertainty related to galaxy mergers.
If a galaxy merger is gas-rich, \citet{del-lagos2018} argue that the specific angular momentum of stars typically increases. If the same applies to star clusters, this would lead to longer dynamical friction timescales and, consequently, a reduced central BH population. In contrast, in the dissipationless major galaxy mergers of \citet{hilz2012a}, the specific energy distribution of particles broadens. This would result in a bimodal effect: the innermost star clusters would sink more rapidly toward the galaxy center, while the inspiral of the outermost ones would be delayed or even suppressed if their specific energy becomes positive.

Our galaxy sample includes both disk- and bulge-dominated galaxies. Therefore, both effects (i.e., increase and decrease of the dynamical friction timescale) likely affect our results.
The mixture of galaxy types will ultimately diffuse clear trends in the resulting BH mass function. We expect that any variation to be smaller or at most of the same order as those introduced by our assumption on the initial stellar density of star clusters.

In summary, in dissipationless major mergers some massive star clusters that host IMBH candidates may gain more specific energy (become more bound and sink towards their host galaxy's center) and may reach the nucleus within a Hubble time.
This effect could extend the mass distribution of BHs towards $\geq 10^{4} \, \textrm{M}_{\odot}$ and result in a direct IMBH seeding scenario for galaxy nuclei.
We explore this mechanism in a separate paper.

\subsubsection{Stellar collisions}

Stellar collisions represent an additional formation scenario for IMBHs in dynamically-active clusters \citep{kremer2020,gonzalesprieto2022, arcasedda2023d2}. This mechanism is most effective in young star clusters \citep{mapelli2016,dicarlo2019}, where short relaxation times drive mass segregation in the very first Myr. Recent studies \citep{fujii2024,vergara2025,rantala2026} indicate that stellar collisions can produce IMBHs $\gtrsim 10^3 \, \msun$ even at $Z \sim 0.02$. 

In the following, we discuss the possible impact of stellar collisions on our results. Including this additional physical process goes beyond the scope of this work, and requires a dedicated model for the interplay between stellar evolution and cluster dynamics. This process is also affected by several uncertainties, related to mass loss and chemical mixing during and after stellar collisions \citep{ballone2022}. 

Recently,  \cite{rantala2026} showed that the maximum IMBH mass produced via stellar collisions increases with the initial cluster half-mass density, exceeding $5000 \, \msun$ in the densest clusters ($\Sigma_{\mathrm{h}} \gtrsim 10^{5} \, \msun \, \mathrm{pc^{-2}}$). The authors consider star clusters with initial mass $10^4-10^6 \,\msun$, and core collapse timescales of the order of massive stars lifetimes, $t_{\mathrm{cc}} \sim 10 \, \mathrm{Myr}$. 

Migrating clusters span across a similar density range, but exhibit core-collapse timescales that are at least an order of magnitude longer ($t_{\mathrm{cc}} \gtrsim 300 \, \mathrm{Myr}$). As shown in \cite{mestichelli2026}, stellar collisions cannot produce runaway sequences in clusters with such relaxation times, but only very limited chains ($\leq 2$), primarily involving primordial binaries. As a result, they are expected to form IMBHs with $m_{\mathrm{BH}} \lesssim 400 \, \msun$, comparable to those from hierarchical mergers in our fiducial model.

\section{Conclusions} \label{sec:concusions}
We investigated the process of IMBH ($m_{\mathrm{BH}} \geq 100 \, \msun$) seeding galactic nuclei from star cluster migration. We introduced \texttt{inSpyral}, a new population-synthesis model that integrates star cluster progressive dissolution and orbital decay in arbitrary galactic potentials, while accounting for BH core dynamics. We initialized the galactic potentials and cluster populations from the galaxy formation model \textit{L-Galaxies 2020} \citep{hoyer2025}. In the following, we report our main results and conclusions.

The most massive star clusters ($M_{\mathrm{sc, 0}} > 5 \times 10^{6} \, \msun$) efficiently migrate to the galactic center as early as $z\sim8$. The migration timescales range from $\sim 10^2 \, \mathrm{Myr}$ to the Hubble time, depending on the initial cluster galactocentric distance. The fraction of migrating clusters approaches unity for the most massive clusters that form in galaxies with $M_{\mathrm{\star, gal}} \leq 10^{10} \, \msun$, where the nucleation peak resides, and drops to zero for $M_{\mathrm{\star, gal}} \geq 10^{11} \, \msun$.

Star cluster migration seeds galactic nuclei with IMBHs from BH mergers as early as $z \sim 6$. At $z=0$, the BH mass distribution extends up to $m_{\mathrm{BH}} = 300 \, \msun$, a factor of five higher than the maximum BH mass produced by in situ formation in the same galaxies.
The IMBH formation efficiency is $\eta_{\mathrm{BH}} = 2 \times 10^{-6} \, \msun^{-1}$ for $M_{\mathrm{\star, gal}} < 10^{10} \, \msun$. 

The formation and migration of IMBHs via hierarchical mergers is regulated by the initial cluster size. 
If massive clusters form with sub-parsec scale radii ($\lesssim 0.5 \, \mathrm{pc}$), the growth of massive BH seeds occurs before migration, thanks to the extremely high escape velocities ($\gtrsim 500 \, \mathrm{km \, s^{-1}}$) and short dynamical timescales. As a result, IMBHs above $10^4 \, \msun$ can form within their parent cluster and then populate galactic nuclei.
If the initial cluster size exceeds the parsec scale, hierarchical mergers are suppressed at the second or third generation, which represent the main bottleneck for the BH growth via mergers.

Gravitational-wave events like GW190521, GW241011, and GW24110 are well reproduced by second-generation BH remnants experiencing relatively small relativistic kicks ($< 100 \, \mathrm{km \, s^{-1}}$). GW231123 is consistent with BH mergers between a third-generation primary and a second-generation secondary, which occur only in massive clusters with $M_{\mathrm{sc,0}} > 2 \times 10^{6} \, \msun$.

In summary, our results indicate that star cluster migration can populate galactic nuclei with IMBHs formed via BH mergers only, independently of uncertainties in cluster formation and BH physics. This scenario produces a maximum BH mass that is a factor of five higher than the upper limit from in situ formation. The maximum BH mass, however, potentially exceeds $10^{4} \, \msun$ if the typical cluster initial size is less than $ 0.5 \, \mathrm{pc}$. Future \textit{JWST} observations of compact proto-star clusters will serve as crucial testbeds to constrain cluster birth properties and, consequently, their role in IMBH seeding in galactic nuclei.

\begin{acknowledgements}

The authors are grateful to the MPIA Galactic Nuclei group and Michela Mapelli for useful discussions and suggestions. ST thanks Alessandra Mastrobuono Battisti, Giuliano Iorio, Alessandro Alberto Trani, Erika Korb, Benedetta Mestichelli and Dani Mar\'in Pina for helpful discussions.

ST acknowledges financial support from the Alexander von Humboldt Foundation for the Humboldt Research Fellowship.

MAS acknowledges funding from the European Union’s Horizon 2020 research and innovation programme under the Marie Skłodowska-Curie grant agreement No.~101025436 (project GRACE-BH), from the MERAC Foundation through the 2023 MERAC prize. MAS acknowledges the ACME project, which has received funding from the European Union's Horizon Europe Research and Innovation programme under Grant Agreement No.~101131928. 

We draw our initial black hole distributions from the \textsc{sevn} catalogs by \cite{iorio2023}, publicly available at \href{https://zenodo.org/records/7794546}{this link}. The \textsc{sevn} code is available at \href{https://gitlab.com/sevncodes/sevn}{this link}. The \textsc{parsec} stellar tracks are publicly available at \href{http://stev.oapd.inaf.it/PARSEC/}{this link} \citep{costa2025}.
The galactic potential and initial cluster populations are based on \textit{L-Galaxies 2020} (publicly available at \href{https://github.com/LGalaxiesPublicRelease/LGalaxies2020_PublicRepository}{this link}, \citealp{henriques2020}) catalogs by \cite{hoyer2025}. 

The galactic potential and orbital integration is modeled with \texttt{galpy} (\citealp{bovy2015}, \href{https://pypi.org/project/galpy/}{link}). We integrate the GW orbital decay with the \textsc{pyblack} package (\citealp{iorio2023}, \href{https://gitlab.com/iogiul/pyblack}{link}). We evaluate the BH remnant properties using the \texttt{surfinBH} models \citep{varma2019,boschini2023},
available at \href{https://pypi.org/project/surfinBH}{this link}.
Our model for cluster evolution draws from \textsc{clusterBH} \citep{antonini2020}, publicly available at this \href{https://github.com/mgieles/clusterbh}{link} (for a more recent version see \href{https://github.com/cBHBd/cBHBd}{here}, \citealp{fronimos2026}).
This research made use of \textsc{NumPy} \citep{Harris20}, \textsc{Polars} \citep{Polars}, \textsc{SciPy} \citep{SciPy2020} and \textsc{Matplotlib} \citep{Hunter2007}.
The development of \textsc{sevn} was enabled by M. Mapelli’s ERC Consolidator grant DEMOBLACK under contract no. 770017. 
The data underlying this article will be shared upon reasonable request to the corresponding author.

\end{acknowledgements}

\bibliographystyle{aa}
\bibliography{references}

\appendix

\section{BH dynamics in the star cluster core} \label{app:bh_bal_ev}
In the cluster core, BBHs can undergo binary-single (BBH-BH) and binary-binary (BBH-BBH) interactions. 
In the following, we provide details for our model of BBH-BH (Appendix \ref{app:bbh_bh_int}) and BBH-BBH (Appendix \ref{app:bbh_bbh_int}) interactions.

\subsection{BBH-BH interactions} \label{app:bbh_bh_int}
We model BBH-BH interactions as a sequence of resonant intermediate states \citep{samsing2014,samsing2018}. The number of intermediate states depends on the mass ratios of the interaction as \citep{randoforastier2025}:
    \begin{eqnarray}
    N_{\mathrm{IMS}} = \left( 1 + q_3 + \frac{q_3}{q} \right)^{\gamma - 1}, \; \; q_3 < q \\
    N_{\mathrm{IMS}} = \left( 1 + q + \frac{q}{q_3} \right)^{\gamma - 1}, \; \; q_3 \geq q, 
    \end{eqnarray}
where $q_3=m_3/m_1$ and $\gamma=3.75$ \citep{randoforastier2025}. For each intermediate state, we generate the eccentricity from a thermal distribution and calculate the GW energy loss due to the passage at pericenter \citep{hansen1972, samsing2014}:
    \begin{equation} \label{eq:gw_loss}
        \Delta E_{\mathrm{GW}} = \frac{85 \pi}{12 \sqrt{2}} \frac{G^{7/2} \, m_1^2  \, m_2^2  \, (m_1+m_2)^{1/2} }{c^5 \, r_{\mathrm{p}}^{7/2}},
    \end{equation}
where $r_{\mathrm{p}}=a \, (1-e)$. If $\Delta E_{\mathrm{GW}}$ exceeds the binary binding energy, the BBH undergoes a GW capture \citep{samsing2018}. 

If the BBH survives the resonant phase, the interaction leads to a fly-by or an exchange.
We evaluate the probability of each scattering outcome from cross-sections derived from the BBH-BH scattering experiments in \cite{sigurdsson1995}, which span across a wide range of mass ratios and velocity regimes. First, we evaluate the critical velocity of the interaction:
\begin{equation}
    v_{\mathrm{c}} = \sqrt{G \, \frac{m_1 \, m_2}{m_3} \frac{m_1+m_2+m_3}{m_1+m_2} \frac{1}{a}}.
\end{equation}
We then select the relative outcome cross-sections in the appropriate $v_{\mathrm{\infty}}/v_{\mathrm{c}} \approx \sigma_{\mathrm{BH}}/v_{\mathrm{c}}$ regime. In our models, we generally have $\sigma_{\mathrm{BH}}/v_{\mathrm{c}} \lesssim 0.1$, implying that dynamical interactions occur in the gravitational focusing regime. 
We model the outcomes of the BBH–BH interaction as follows.

\paragraph{\textbf{Fly-by.}} The BBH binding energy variation in a fly-by is \citep{randoforastier2025}:
    \begin{equation}
        \Delta E_{\mathrm{BBH}} = \Delta E_{\mathrm{BBH, 0}} \left[ 1 - \exp\left( 1 - A \frac{m_3}{m_1+m_2} \right) \right],
    \end{equation}
where $A=7.0$. The parameter $\Delta E_{\mathrm{BBH, 0}} =0.2$ is the energy variation for an equal-mass scattering \citep{samsing2018}, and corresponds to a semi-major axis shrinking by $\delta_{\mathrm{bs}} = a_{\mathrm{f}} / a_{\mathrm{i}} = 0.83$, where $a_{\mathrm{i}}$ ($a_{\mathrm{f}}$) is the semi-major axis before (after) the interaction. More asymmetric BBH-BH scatterings lead to smaller energy variations. 
We sample the new BBH eccentricity from a thermal distribution.

\paragraph{\textbf{Exchange.}} We sample the final semi-major axis from the distribution (Eq.~4.2 in \citealp{sigurdsson1995}):
\begin{equation} \label{eq:a_exchange}
    p(a_{\mathrm{f}}/a_{\mathrm{0}}) \sim \frac{a_{\mathrm{f}} \, e^{(a_{\mathrm{f}}-a_{\mathrm{0}})}}{e^{3(\sigma_{\mathrm{BH}}/v_{\mathrm{c}})(a_{\mathrm{f}}-a_{\mathrm{0}})}-1},
\end{equation}
where $a_{\mathrm{0}} \approx a_{\mathrm{i}} \left[ 1 + (\sigma_{\mathrm{BH}}/v_{\mathrm{c}})^2 \right](m_3/m_{\mathrm{ex}})$ and  $m_{\mathrm{ex}}$ is the BH component that is exchanged. In general, exchanges lead to BBH hardening, with $a_{\mathrm{f}} \lesssim a_{\mathrm{i}} \, (m_3/m_{\mathrm{ex}})$. We draw the BBH eccentricity after an exchange from a thermal distribution.

\subsection{BBH-BBH interactions} \label{app:bbh_bbh_int}
In a BBH-BBH interaction, we estimate the probability of a GW capture merger during the resonant phase from the fitting formula in  \cite{marinpina2025}:
    \begin{equation}
        p_{\mathrm{merge}} = 3.4\times 10^{-2} \left( \frac{ \langle m \rangle}{20 \, \msun} \right)^{5/7} \left( \frac{ a_{\mathrm{min}}}{0.1 \, \mathrm{AU}} \right)^{-5/7} \left[ 1+ \left( \frac{\alpha}{8.6} \right)^2 \right]^{-0.83},
    \end{equation}
where $\langle m \rangle$ is the mean BBH component mass and $\alpha = a_{\mathrm{max}} / a_{\mathrm{min}}$. If the BBH merges, we sample the eccentricity such that the gravitational-wave energy loss (Eq.~\ref{eq:gw_loss}) exceeds the BBH binding energy. In this way, we assure that the GW energy loss due to the passage at the pericenter is large enough to cause the merger. 

If the BBH survives the resonant phase, we sample the outcome of the BBH-BBH interaction based on cross-sections from the scattering experiments by \cite{antognini2016}. The BBH-BBH interaction outcomes mainly depend on the velocity regime and the semi-major axis ratio, with much milder dependence on mass ratios and eccentricity \citep{antognini2016}. We define the critical velocity of the BBH-BBH interaction as:
\begin{equation}
    v_{\mathrm{c,bb}} = \sqrt{\frac{G (m_{12}+m_{34})}{m_{12} \, m_{34}}} \left( \frac{m_1 \, m_2}{a_1} + \frac{m_3 \, m_4}{a_2} \right),
\end{equation}
where $m_{12}=m_{1}+m_{2}$ and $m_{34}=m_{3}+m_{4}$. We derive $\sigma_{\mathrm{BH}} / v_{\mathrm{c,bb}}$ to select the proper velocity regime. Also in this case, the interaction generally takes place in the gravitational focusing regime, with $\sigma_{\mathrm{BH}}/v_{\mathrm{c, bb}} \lesssim 0.1$. 
Then, we calculate the cross-section in the proper velocity regime and for the given semi-major axis ratio. Here, the possible outcomes are fly-by, exchange, ionization, and triple formation. 

\paragraph{\textbf{Fly-by.}} In a BBH-BBH interaction, fly-bys generally imply a negligible energy variation (\citealp{bacon1996}, Torniamenti et al. in prep.). In this case, we assume that the two BBHs do not undergo any hardening, and we sample the new eccentricities from a thermal distribution. 
   
\paragraph{\textbf{Exchange.}} We assume that the primary mass of the BBH interloper is exchanged for the secondary BH of the interacting BBH. Then, we sample the new semi-major axis in the same way as binary-single interactions \citep{sigurdsson1995}, following Eq.~\ref{eq:a_exchange}. We sample the eccentricities for the exchanged BBHs from a thermal distribution.
    
\paragraph{\textbf{Ionization.}} During a ionization, the softer BBH is disrupted, and the semi-major axis of the surviving BBH hardens by a factor \citep{zevin2019}:
    \begin{equation}
        \delta_{\mathrm{bb}} = \frac{24 \, \delta_{\mathrm{bs}}}{51 - 3 \, \delta_{\mathrm{bs}}}.
    \end{equation}
Following \cite{antognini2016}, the ionization can be preceded by an exchange (\textbf{Exchange+Ionization}). In this case, we first perform an exchange and then ionize the resulting softer BBH. We sample the eccentricity of the surviving BBH from a thermal distribution.

\paragraph{\textbf{Triple formation.}} 
When a stable triple forms, the softer BBH is ionized, and its primary BH becomes the outer member of the triple. The internal BBH semi-major axis hardens by an average factor $\delta_{\mathrm{trip}} = 0.8$, consistently with BBH-BBH scatterings in Monte Carlo simulations \citep{arcasedda2021}. We consider a semi-major axis ratio $a_{\mathrm{out}}/a_{\mathrm{in}} = 100$ (\citealp{arcasedda2021}, Torniamenti et al. in prep.), and we draw the mutual inclination ($i$) from a uniform distribution \citep{trani2022,hayashi2023}. We sample the outer eccentricity from a thermal distribution, provided that the triple system fulfills the stability criterion \citep{mardling2001}:
\begin{equation}
    \frac{a_{\mathrm{out}}}{a_{\mathrm{in}}} = \frac{2.8}{1-e_{\mathrm{out}}} \left[ \left( 1 + q_{\mathrm{out}} \right) \, \frac{1+ e_{\mathrm{out}}}{\sqrt{1-e_{\mathrm{out}}}} \right]^{2/5} \left( 1 - 0.3 \frac{i}{\pi} \right),
    \end{equation}
where $q_{\mathrm{out}}= m_3 / (m_1+m_2)$. This requirement quenches the formation of triple systems with $e_{\mathrm{out}}\approx 1$ (\citealp{marinpina2025}, Torniamenti et al. in prep.).

 In a stable triple system, the inner BBH can merge for GW emission triggered by ZLK oscillations \citep{vonzeipel1910,lidov1962,kozai1962}, producing mergers that are eccentric in the LIGO-Virgo-KAGRA band.
 We select the possible candidates for ZLK mergers as those with $|\cos{(i)}|<3/5$, for which the inner binary eccentricity can reach a maximum value \citep{naoz2016}:
    \begin{equation}
        e_{\mathrm{max}} = \sqrt{1- \frac{5}{3} \cos^2{(i)}}.
    \end{equation}
    We then calculate the timescale for ZLK oscillations \citep{arcasedda2021}:
    \begin{equation}
    t_{\mathrm{ZLK}} = \frac{8}{15 \, \pi} \left( 1+ \frac{m_1+m_2}{m_1+m_2+m_3} \right) \left( \frac{P^2_{\mathrm{out}}}{P_{\mathrm{in}}} \right) \left( 1 - e^2_{\mathrm{out}}\right)^{3/2},
    \end{equation}
    where $P_{\mathrm{in}}$ ($P_{\mathrm{out}}$) is the inner (outer) period of the triple system. During a ZLK cycle, we determine whether a merger occurs from Eq.~\ref{eq:gw_loss}, which requires the eccentricity to fulfill:
    \begin{equation}
        e_{\mathrm{GW,ZLK}} = 1 - \left(\frac{85\pi} {12 \sqrt{2}} \right)^{2/7} \left[ \frac{G^5 \, m^2_1 \, m^2_2 \, (m_1+m_2)}{ a^5_{\mathrm{in}} \, c^5} \right]^{1/7}
    \end{equation}
A stable triple undergoes a merger due to ZLK oscillations if $e_{\mathrm{\mathrm{max}}} \geq e_{\mathrm{GW,ZLK}}$ and the time for the next interaction (or the Hubble time, if the triple system is ejected) is larger than $t_{\mathrm{ZLK}}$. Otherwise, we assume that it is ionized before the next interaction.

We find that that $50\%$ BBH-BBH interactions lead to BBH conservation, $\sim 30\%$ to BBH ionization, and  $\sim 20\%$ to triple formation, when considering a semi-major axis ratio distribution ranging from $1$ to $10^5$, as suggested by BBH-BBH interactions in Monte Carlo simulations \citep{arcasedda2021,marinpina2025}. These fractions are broadly consistent with the results of BBH-BBH scatterings in the gravitational focusing regime \citep{arcasedda2021,marinpina2025}.

\section{Stellar mass loss and relaxation} \label{app:internal_evolution}

\begin{figure*}
    \centering
    \includegraphics[width=\hsize]{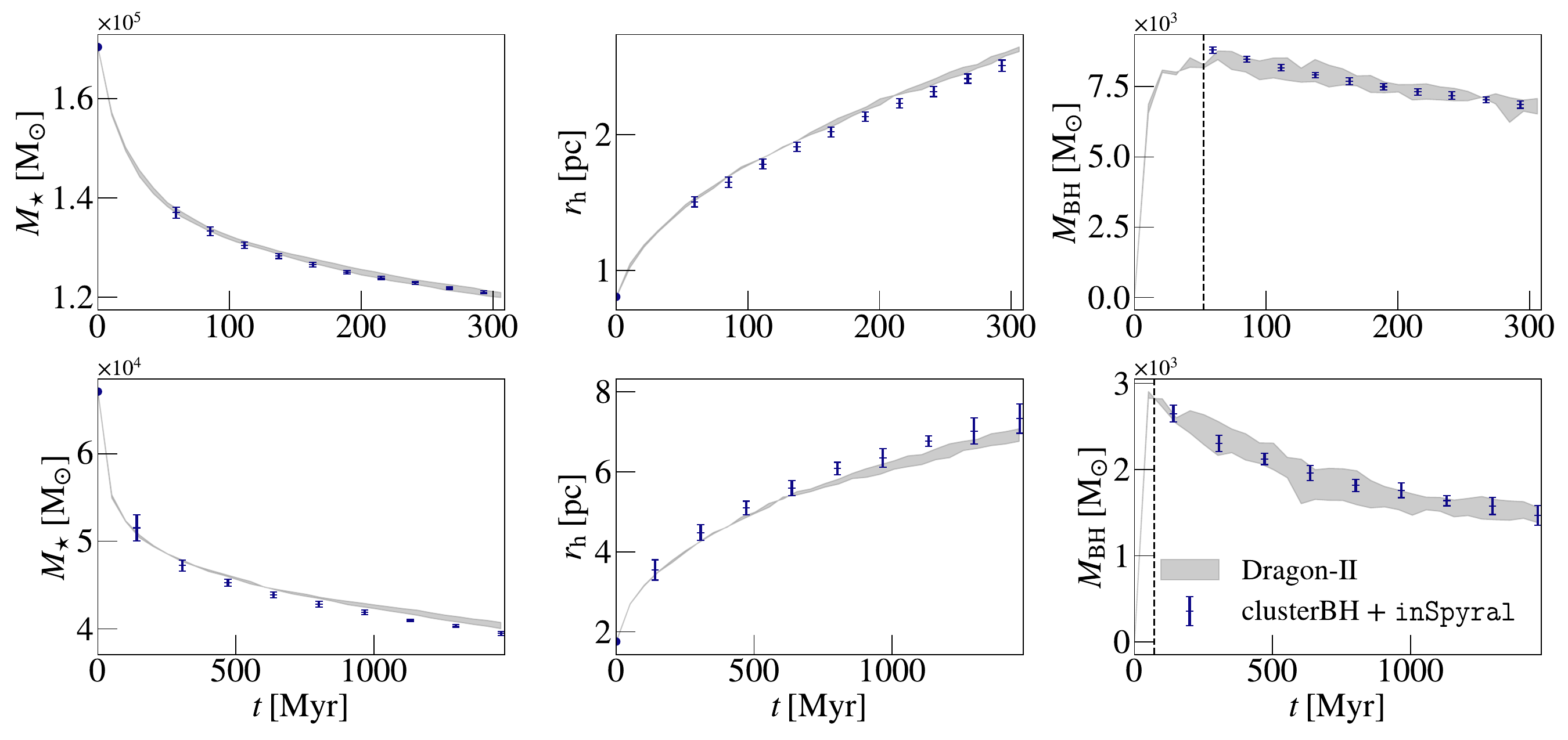}
    \caption{Evolution of the total stellar mass (left), half-mass radius (center) and total BH mass (right) in our model (blue) and the Dragon-II $N-$body models (grey). The purple region shows the area bracketed by the evolution of two $N-$body realizations with the same initial conditions. The black dashed line indicates the time of star cluster core collapse. We model stellar mass loss with \textsc{clusterBH} \citep{antonini2020}, we calibrate cluster expansion to reproduce the \textsc{Dragon-II} simulations \citep{arcasedda2023d1}, and we integrate the total BH mass loss based on our model for BBH-(B)BH scatterings in a state of balance evolution.     
    }
    \label{fig:scatter_comparison}
\end{figure*}

We model star cluster evolution (stellar mass loss and relaxation) with \textsc{clusterBH} 
\citep{antonini2020}. The total stellar mass decreases because of stellar winds and supernova explosions as:
\begin{equation}
\dot{M}_{\star,\mathrm{sev}}  =  
- \nu \, \dfrac{M_{\star}}{t},
\end{equation}
where $t_{\mathrm{sev}}= 3 \, \mathrm{Myr}$ incorporates the change in the stellar-mass loss slope due to the first supernova explosions and $\nu=0.07$ is set to reproduce the evolution of direct Dragon-II  $N-$body simulations (see \citealp{torniamenti2024}). We parametrize the consequent cluster expansion as: 
\begin{equation}
\dot{r}_{\mathrm{h,sev}}   =  - \alpha_{\mathrm{ev}} \frac{\dot{M}_{\star,\mathrm{sev}}}{ {M_{\mathrm{sc}}}}  r^{\beta_{\mathrm{ev}}}_{\mathrm{h}},
\end{equation}
where we calibrate $\alpha_{\mathrm{ev}}$ and $\beta_{\mathrm{ev}}$ to reproduce the trend in the Dragon-II simulations \citep{arcasedda2023d2}. We include the parameter $\beta_{\mathrm{ev}}$ to account for cluster expansion stronger than the adiabatic case ($\beta_{\mathrm{ev}}=1$). This can occur, e.g., due to primordial binary heating before the onset of core collapse \citep{trenti2006,trenti2007}.

\subsection{Relaxation} \label{sec:relaxation}
In our model, $M_{\mathrm{BH}}$ decreases as BHs are ejected from the star cluster core by the dynamical recoils in BBH-(B)BH interactions and relativistic kicks in BBH mergers. In a state of balanced evolution, the time between two interactions can be calculated as the ratio between the energy released by the BBH hardening, $\Delta E_{\mathrm{BBH}}$, and the energy flow through the cluster \citep{antonini2020}:
\begin{equation}
    t_{\mathrm{dyn}} = \frac{\Delta E_{\mathrm{BBH}}} {\zeta \, G M^2_{\mathrm{sc}}/r_{\mathrm{h}}} t_{\mathrm{rh}} ,
\end{equation}
where  $\zeta=0.08$ and $t_{\mathrm{rh}}$ is the cluster relaxation time, defined as \citep{spitzer1971}:
\begin{equation}
    t_{\rm{rh}} = 0.138 \sqrt{\frac{M_{\rm{sc}}\,{} r^3_{\rm{h}}}{G}} \frac{1}{\langle m \rangle \,{}\psi \,{}\rm{\ln{\Lambda}}},
\end{equation}
where $\langle m \rangle$ is the mean mass, ${\ln{\Lambda}}=\ln{(0.4 \, N)}$ is the Coulomb logarithm and $N$ is the total number of stars. The parameter $\psi$ depends on the mass function and relaxation state as \citep{spitzer1971}:
\begin{equation}
\psi = \left(m^{3/2}_{\star}\,{}M_{\star}+m_{\rm{BH}}^{3/2}\,{}M_{\rm{BH}}\right) \, \frac{N^{3/2}_{\rm{sc}}}{ M^{5/2}_{\rm{sc}}}.
\end{equation}
The resulting expansion rate is derived under the assumption of virial equilibrium \citep{antonini2020}:
\begin{equation}
    \dot{r}_{\mathrm{h,bal}} = \zeta \,{}\frac{r_{\mathrm{h}}}{t_{\mathrm{rh}}} + 2 \,{}\frac{\dot{M}_{\mathrm{sc}}}{M_{\mathrm{sc}}} \,{}r_{\rm{h}}.
\end{equation} 

\subsection{Comparison with $N-$body models}
We assess the goodness of our cluster evolution model by comparing the stellar and BH mass evolution to the Dragon-II direct $N-$body simulations \citep{arcasedda2023d2,arcasedda2023d1,arcasedda2023d3}. First, we initialize our models to match the initial conditions of the Dragon-II simulations. To this purpose:
\begin{itemize}
    \item We initialize our models with the same initial mass, radius, and metallicity as a set of Dragon-II runs. We consider two models with initial mass $M_{\mathrm{sc,0}}= 1.7\times10^5$ and $6.7\times10^5 \, \msun$ and half-mass radius $r_{\mathrm{h,0}}=0.8$ and $1.75 \, \mathrm{pc}$ respectively. We consider $Z=0.0006$, which is the closest in our \textsc{sevn} catalogs to the Dragon-II metallicity ($Z=0.0005$).
    \item We evolve our models in the same tidal field as the Dragon-II simulations, that is, a point-mass galaxy with $M_{\mathrm{gal}}=1.78\times10^{11} \, \msun$. We consider a circular orbit at galactocentric distance $R_{\mathrm{g}}=13.3 \, \mathrm{kpc}$.
    \item To compensate for the differences in stellar and binary evolution models, we truncate the BH mass distribution to $40.5 \, \msun$, which is the lower edge of the pair-instability mass gap in \cite{arcasedda2023d1}.  
    
\end{itemize}

For each model comparison, we consider two $N-$body runs with the same initial conditions and $10^3$ semi-analytic realizations. Figure \ref{fig:scatter_comparison} shows the comparison between our model and direct $N-$body simulations. 
For $\alpha_{\mathrm{ev}}=2$ and $\beta_{\mathrm{ev}}=1.3$, our model can reproduce the star cluster expansion in the $N-$body simulations, which is slightly stronger than an adiabatic expansion before core collapse. In conclusion, our semi-analytic modeling based on successive BBH-(B)BH encounters in a state of balance evolution, can reproduce the evolution of the global properties of star clusters in the Dragon-II simulations. 

\section{Original BBH mergers} \label{sec:original_mergers}

\begin{figure*}
    \centering
    \includegraphics[width=\hsize]{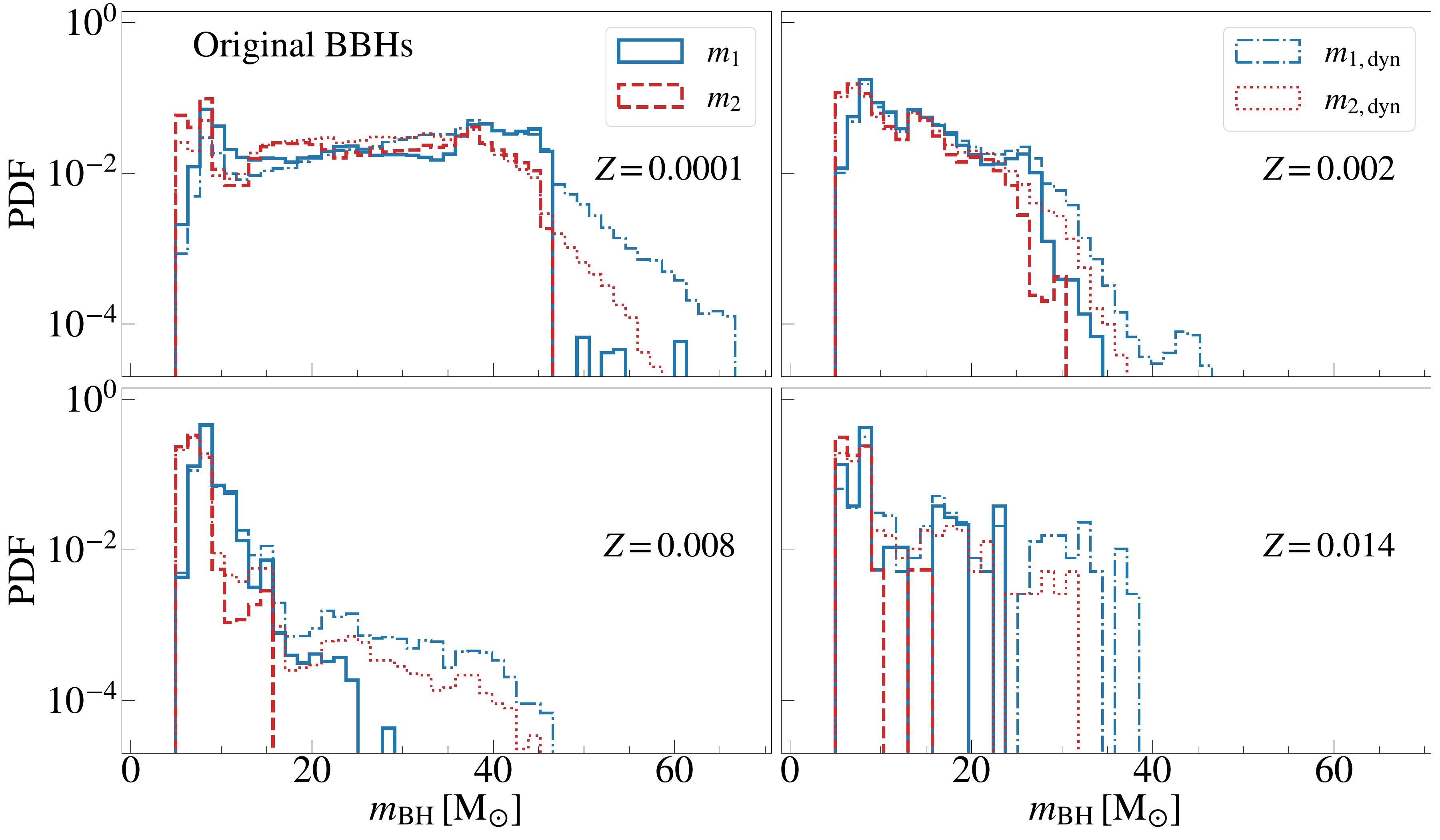}
    \caption{Primary (blue, solid) and secondary (red, dashed) mass distributions of original BBH mergers, for different metallicities: $Z=0.0002$, $0.002$, $Z=0.008$, $Z=0.014$. The dashdot (dotted) lines represent the primary (secondary) mass distribution including BBHs that are bound since their formation, but undergo dynamical interactions.}
    \label{fig:BBH_original_mergers_hist}
\end{figure*}

Figure \ref{fig:BBH_original_mergers_hist} shows the mass distribution of original BBH mergers in our models, for different metallicities. Original BBH mergers exhibit a mass distribution that peaks at low BH masses ($m_{\mathrm{BH}}\sim 10 \, \msun$). These systems are either ejected by natal kicks or retained within their parent star cluster. In the latter case, their semi-major axis at formation is generally much smaller than the typical BH separation because of previous hardening by stable mass transfer and/or common-envelope evolution (see, e.g., \citealp{torniamenti2022}). As a result, their interaction rate with surrounding BHs is strongly suppressed (see Eq.~\ref{eq:int_rate}), and they eventually merge without experiencing significant dynamical interaction. Although these BBHs remain within the star cluster throughout their evolution, their formation is driven by binary evolution, as in the isolated case (see, e.g., \citealp{iorio2023}).

Our star cluster models also host a population of initially wide BBHs from original binaries that merge after interacting with the surrounding BHs (see, e.g., $N-$body simulations from \citealp{barber2025}). 
These BBHs form close to the hard-soft boundary and do not undergo exchanges, because their BH components are already among the most massive within the cluster. Their relatively wide initial separations prevent significant mass-transfer episodes (e.g., stable mass transfer or common envelope), allowing for the formation of original BBHs with masses $m_{\mathrm{BH}} > 40 \, \msun$. Since these BBHs undergo dynamical interactions with the BH core, we will include them in the dynamical BBH population.

\end{document}